\documentclass[11pt,a4paper]{article}
\usepackage{jheppub}
\usepackage{multirow}
\newcommand{\be}{\begin{equation}}
\newcommand{\ee}{\end{equation}}
\newcommand{\ba}{\begin{eqnarray}}
\newcommand{\ea}{\end{eqnarray}}
\def\bs{\begin{subequations}}
\def\es{\end{subequations}}
\def\a{\alpha}
\def\b{\beta}

\def\cM{{\cal M}}

\def\cR{{\cal R}}
\def\cS{{\cal S}}

\def\p{\partial}

\newcommand{\m}{\mu}
\newcommand{\n}{\nu}

\newcommand{\Rb}{\bar{R}}
\newcommand{\Ebb}{{\bf E}}
\newcommand{\ebb}{{\bf e}}

\usepackage{MyJHEPStyle}
\newcommand{\gb}{\bar{g}}
\title{RG flows of Quantum Einstein Gravity \\
on maximally symmetric spaces}
\author[a]{Maximilian Demmel,}
\author[b]{Frank Saueressig,}
\author[b]{Omar Zanusso}
\affiliation[a]{
PRISMA Cluster of Excellence \& Institute of Physics (THEP), \\
University of Mainz, Staudingerweg 7, D-55099 Mainz, Germany
}
\affiliation[b]{
Institute for Mathematics, Astrophysics and Particle Physics (IMAPP),\\
Radboud University Nijmegen, Heyendaalseweg 135, 6525 AJ Nijmegen, The Netherlands
}
\emailAdd{demmel@thep.physik.uni-mainz.de}
\emailAdd{f.saueressig@science.ru.nl}
\emailAdd{o.zanusso@science.ru.nl}
\abstract{
We use the Wetterich-equation to study the renormalization group flow of $f(R)$-gravity in a three-dimensional, conformally reduced setting. Building on the exact heat kernel for maximally symmetric spaces, we obtain a partial differential equation which captures the scale-dependence of $f(R)$ for positive and, for the first time, negative scalar curvature. The effects of different background topologies are studied in detail and it is shown that they affect the gravitational RG flow in a way that is not visible in finite-dimensional truncations. Thus, while featuring local background independence, the functional renormalization group equation is sensitive to the topological properties of the background. The detailed analytical and numerical analysis of the partial differential equation reveals two globally well-defined fixed functionals with at most a finite number of relevant deformations. Their properties are remarkably similar to two of the fixed points identified within the $R^2$-truncation of full Quantum Einstein Gravity. As a byproduct, we obtain a nice illustration of how the functional renormalization group realizes the ``integrating out'' of fluctuation modes on the three-sphere.
}
\begin{document}
\maketitle
%
\section{Introduction}
\label{sect.1}

It is very well known that general relativity is perturbatively non-renormalizable
because at every loop order new operators require renormalization
\cite{tHooft:1974bx,Goroff:1985sz,vandeVen:1991gw}.
Thus, from a perturbative quantum field theory perspective,
general relativity can only have the status of an effective field theory
which is valid up to a certain high, but finite, UV scale $\Lambda$
that can be argued to be of the order of the Planck's mass.
The perturbative non-renormalizability of general relativity
is directly related to the lack of asymptotic freedom
in the renormalization group (RG) flow of its couplings and, in particular,
of Newton's constant.

An alternative scenario has been proposed by Weinberg \cite{Weinberg:1980gg} (see also \cite{Weinproc1}),
who noticed that gravity may instead be asymptotically safe.
In this scenario, reviewed in many references \cite{Niedermaier:2006wt,Reuter:2007rv,robrev,Litim:2008tt,Reuter:2012id},
the Newton's constant
runs towards a non-Gaussian fixed point (NGFP) in the UV.
The predictivity of an asymptotically safe theory is ensured if the
NGFP comes with a finite dimensional critical surface $\cS_{\rm UV}$,
which is spanned by the set of RG trajectories
that flow to the NGFP as the RG scale approaches infinity.
An asymptotically safe theory of gravity, also known as Quantum Einstein Gravity (QEG),
is therefore both consistent,
because in the UV all observables such as scattering amplitudes approach a finite value in units of the RG scale,
and predictive, because Nature's realized gravity can be located on the critical surface by a finite
number of experiments. As a simple corollary, the lower is the dimension of $\cS_{\rm UV}$,
the higher is the predictive power of the asymptotically safe theory
as the number of required measurements equates the dimensionality of the surface.

Investigating the existence of a NGFP in the RG flow of a theory
requires a non-perturbative tool,
since the validity of perturbation theory is by definition restricted
to a neighborhood of the Gaussian fixed point (GFP)
where all interactions vanish
and the scaling dimensions of the operators match the ones obtained from classical power counting.
The exact renormalization group, also known as functional renormalization group (FRG),
offers such a non-perturbative tool.
The fundamental ingredient of the FRG method is an effective average action
$\Gamma_k$
that is coarse-grained at a reference RG scale $k$
\cite{Wetterich:1992yh}.
The flow of $\Gamma_k$ satisfies the exact
functional renormalization group equation (FRGE)
\be\label{FRGE}
\p_t \Gamma_k[\Phi, \bar{\Phi}] = \frac{1}{2}  {\rm STr}  \left[ \left( \Gamma_k^{(2)} + \cR_k \right)^{-1} \, \p_t \cR_k \right] \, .
\ee
Here $t \equiv \ln(k)$ is the renormalization group time, $\Phi$ denotes the set of fluctuation fields and $\bar{\Phi}$ their background value. $\Gamma_k^{(2)}$ is the second variation
of $\Gamma_k$ with respect to $\Phi$ at fixed $\bar{\Phi}$, $\cR_k$ is a IR cutoff, providing a mass term for fluctuations with momenta $p^2 \le k^2$,
and the STr contains an integral over loop momenta together with a sum over all fluctuation fields. The interplay of $\cR_k$ appearing in the numerator and
denominator renders the r.h.s.\ of the equation finite and peaked at momenta $p^2 \approx k^2$. The RG flow of $\Gamma_k$ is thus driven by integrating out quantum fluctuations
close to the reference scale $k$. In this sense $\Gamma_k$ constitutes a Wilsonian effective action that provides an effective description for the physics at a typical momentum scale $k$.
The flow equation  \eqref{FRGE} can be understood as a system of infinitely many coupled
partial differential equations among the irreducible vertices of the effective theory.
For this reason, it is extremely difficult to find exact solutions.
In absence of a guiding principle such as the expansion in powers of the marginal couplings that underlies perturbation theory \cite{Codello:2013bra},
some kind of approximation must be employed to the flow
to make it tractable by either analytical or numerical approximations.
Most of the FRG studies of gravity approximated the flow
by truncating $\Gamma_k$ to a finite subset of the space of operators.
Starting from the seminal works \cite{Reuter:1996cp,Dou:1997fg,Souma:1999at},
the gravitational RG flow has been successively projected onto  subspaces of increasing complexity.
In particular the existence of a NGFP has been established in the
Einstein-Hilbert truncation \cite{Lauscher:2001ya,Reuter:2001ag,Nagy:2012rn,Donkin:2012ud,Christiansen:2012rx},
the $R^2$ truncation \cite{Lauscher:2001rz,Lauscher:2002sq,Rechenberger:2012pm},
$f(R)$ truncations \cite{Codello:2007bd,Machado:2007ea,Codello:2008vh,Bonanno:2010bt,Falls:2013bv,Rahmede:2011zz},
and truncations including a Weyl-squared term \cite{Benedetti:2009rx,Benedetti:2009gn}.
Moreover, refs.\ \cite{Reuter:2001ag,Fischer:2006at} studied the properties of the NGFP in spacetime dimension more than four, while the quantum effects in the ghost sector have been investigated in \cite{Eichhorn:2009ah}. The investigation of ``bi-metric'' truncations has been initiated in \cite{Manrique:2010am} and flows including boundary terms relevant for black hole physics have been considered in \cite{Becker:2012js}. The signature dependence of the NGFP was investigated in \cite{Manrique:2011jc} and a computer based algorithm for evaluating the flow equations was proposed in \cite{Benedetti:2010nr,Saueressig:2011vn}. Finally, a physical explanation for asymptotic safety based on paramagnetic dominance has been advocated in \cite{Nink:2012vd}.
The finite-dimensional truncations offered substantial evidence supporting the existence of the NGFP
for various choices of the operators and of the coarse-graining schemes.
As a general argument, a fundamental quantum theory of gravity based on a NGFP
is very likely predictive \cite{Reuter:2002kd,Codello:2010mj,Satz:2010uu,Codello:2011js}.
In fact, if this was not the case, quantum effects would have
to convert an infinite number of classically irrelevant local operators into relevant ones.
Such a failure of predictivity requires the presence of large anomalous dimensions.
While this is a very unlikely occurrence \cite{Codello:2013fpa}, it cannot be excluded a priori \cite{Vacca:2010mj}.


Finite dimensional truncations are not capable of testing
whether the critical surface of a NGFP is actually finite dimensional.
Since predictivity is a core motivation for pursuing the gravitational asymptotic
safety program, it is then natural
to seek for alternative approximations that are capable of providing
a solid argument in favor of a finite-dimensional critical surface.
The simplest and most natural approximation in these regards is to include
an infinite number of coupling constants in the ansatz for the effective average action.
Identifying the analogue of the NGFP in such a setting
and establishing that the critical surface is finite dimensional,
despite probing an infinite-dimensional space of coupling constants,
would constitute a very strong argument in favor
of the asymptotic safety conjecture.

One promising line of research in this program includes infinitely many operators in the gravitational effective average action by approximating $\Gamma_k$ by a $f(R)$-truncation.
In this case $\Gamma_k$ is truncated to a functional of the scalar curvature $R$
\begin{eqnarray}\label{fRans}
 \Gamma^{\rm grav}_k[g]=\int {\rm d}^{d}x \sqrt{g} f_k(R)\,,
\end{eqnarray}
and supplemented by suitable gauge-fixing and ghost terms.
Here $g_{\mu\nu}$ is the metric and $f_k(R)$ is an arbitrary function
depending on the RG scale $k$.
This ansatz has the advantage that the resulting
partial differential equation (PDE) encoding the scale-dependence of $f_k(R)$
can be constructed using a maximally symmetric background geometry
where all curvature invariants can be expressed in terms of $R$.
The equation can generally be obtained analytically for some choice of the coarse-graining scheme
and its solutions can be studied numerically.
In this setting, fixed point solutions
are $k$-stationary solutions (fixed functions), $f_*(R)$, of the PDE describing the flow of $f_k(R)$.
Obtaining the fixed function solutions is generally a very demanding numerical task,
as we will discuss below.
Obviously, the function $f_k(R)$ contains an amount of information equivalent to infinitely many couplings as can be deduced, for example, by Taylor expanding it in local operators of the form $R^n$.
This program has been applied to both four-dimensional
\cite{Machado:2007ea,Benedetti:2012dx,Dietz:2012ic,Benedetti:2013jk,Dietz:2013sba,Benedetti:2013nya}
and three-dimensional \cite{Demmel:2012ub,Demmel:2013myx} quantum gravity.
Both applications are of interest because gravity is expected to be asymptotically safe in both cases \cite{Ohta:2012vb,Ohta:2013uca}.
In fact, the three-dimensional case offers a simplified environment to develop, test and refine
the needed analytical and numerical tools, that could be later applied to the more physically interesting
four-dimensional case.

For convenient choices of the truncation scheme, the fixed point equation for $f_k(R)$
turns out to be a third order ordinary differential equation (ODE) in $R$
that can be solved numerically.
Also, the ODE admits three singularities in the form of poles located at certain values of the variable $R$.
It is instructive to compare the features of this equation
to an analog equation for the potential of a scalar field obtained in the local potential approximation (LPA)
\cite{Wetterich:1992yh,Morris:1994ie,Morris:1998da,Bridle:2013sra,Canet:2003qd,Litim:2010tt}.
In the LPA the fixed point equation for the potential is of second order and admits only one pole singularity,
making it considerably simpler than the gravitational case.
The simpler structure is thus reflected in some successful achievement
for the corresponding numerical analysis \cite{Codello:2012sc,Codello:2012ec}.

It is interesting to attempt to answer how many global solutions
are expected to the gravitational fixed function program.
The number of the possible solutions of the ODE can be easily estimated
in the following way: a third order ODE has a space of solutions
that is generally characterized by three parameters. However, if we require the solution to exist
for any value of the dimensionless combination $r = R/k^2$,
it is necessary to carefully tune these parameters to get rid of the singularities at finite $r$.
It has been conjectured, and up to now always observed in practice, that the number
of parameters that have to be tuned equates the order of the ODE,
thus indicating that most likely there are only a countable number fixed functional solutions.
If we define the index of the equation as the difference of its order and the number of singularities \cite{Dietz:2012ic},
\begin{equation}\label{indexcounting}
\mbox{index} = \mbox{order of differential equation} - \mbox{number of singularities}
\end{equation}
it is expected both to be zero and to be insensitive of the coarse-graining implementation.
For example, if for a different cutoff the equation is of second order,
it is expected to have only two singularities at finite $R$ \cite{Demmel:2012ub}.

The fixed-functional solution of $f_k(R)$ has already been studied extensively, both perturbatively \cite{Cognola:2005de,Demmel:2013myx} and non-perturbatively in $R$ \cite{Benedetti:2012dx,Demmel:2012ub,Dietz:2012ic,Benedetti:2013jk,Dietz:2013sba,Bridle:2013sra}.
Since we seek for solutions that are finite in units of the scale $k$,
the fixed function of interest comes as a stationary solution for the flow of the dimensionless
$\varphi_k(r) \equiv k^{-d}f_k(k^2 r)$.
For a fixed curvature scalar $R$, the UV, $k\to\infty$, corresponds to $r\to 0$.
This suggests that the UV behavior of the fixed function should be understood
consistently in a small-$r$ polynomial expansion and culminated in some recently successful exploration \cite{Falls:2013bv} where the polynomial expansion of the flow
of $f(R)$ was explored in order to obtain a solution
which is valid perturbatively in the curvature to a very high order.
As we shall see later in this paper, the small-$r$ expansion is however completely insensitive of contributions
that are non-local in the variable $r$. It might be argued,
and observed in practice for the three-dimensional case,
that the expansion may not be the best approximation.
Further, we do not know a priori whether the IR effective action admits a polynomial expansion,
therefore there seems to be a fundamental gap in understanding how a perturbative solution
should relate to a non-perturbative one.
We will be seeking for an implementation of the fixed-functional solution
which goes beyond the simple small-$r$ expansion and at the same time captures fully the non-local
contributions in $r$ of the flow in the maximum range of existence of $r$.

The existence of such a solution is a valuable ingredient in the asymptotic safety scenario,
because it has already been argued \cite{Benedetti:2013jk} that, if a global solution exists,
it will admit a discrete spectrum of deformations and only a finite number of them will be relevant,
in agreement with the requirement of asymptotic safety of a finite dimensional ${\cal S}_{\rm UV}$.

There are still many open questions in the framework of the fixed-functional solutions
and some have only partial answers.
It is still rather unclear how a solution can strictly be global.
Naively, a global solution is a solution that extends in the whole range of the IR cutoff $k$.
For a fixed value of the scalar curvature $R$, this implies that the solution should extend
for all the possible values of the dimensionless scalar curvature
$r \in (-\infty,\infty)$.
Deforming a given geometry from positive to negative $r$ implies a change in the (background) topology,
and therefore a global solution in $r$ will not be global in a topological sense.
Further, it is unclear how the background topology affects the ODE determining possible
fixed functionals. However, due to the topological obstruction, analytic continuations of the solution for positive values of $r$ are not expected to hold for negative values and vice-versa.
A derivation of the flow equation on geometries characterized by negative $R$ has never been carried out without the aid of analytic continuation,
but is a topic that will be carefully addressed in this work.
Thus, despite the effort that was put in both three \cite{Demmel:2012ub} and four dimensions \cite{Benedetti:2012dx,Dietz:2012ic,Benedetti:2013jk,Dietz:2013sba}, nobody has yet managed
to construct a solution that is valid in the whole range $r \in (-\infty,\infty)$.

In this work we attempt to develop a coherent picture concerning the questions discussed above.
On the analytic side it will provide the derivation of a one-parameter family of PDEs
encoding the flow of $f_k(R)$ which covers both the positive and negative curvature domain.
As a novel feature this equation implicitly takes the effect of the topology change into account.
This will be achieved by carefully expressing the flow in terms of the heat kernel \cite{Avramidi:2000bm}
of the Laplace-type operator used as a reference for the coarse-graining procedure.
The implementation will thus use a maximally symmetric background \cite{Camporesi:1990wm},
as the spectrum of Laplace-operators
and the corresponding eigenfunctions are known explicitly \cite{Rubin:1984tc,Camporesi:1994ga}.
We establish that the background topology plays
an important role, leading to non-local contributions in $R$ which were missing from
many previous works in the literature. The resulting flow equation gives 
a transparent interpretation how the FRG integrates out fluctuations
around the background geometry.

In order to avoid being swamped by technical details, we will carry out the computation
for the toy-model of three-dimensional conformally reduced gravity \cite{Reuter:2008wj,Reuter:2008qx,Reuter:2009kq,Machado:2009ph,Bonanno:2012dg,Hooft:2010nc}.
In this case the flow only takes into account fluctuations of the metric that
are proportional to their trace, thus being infinitesimal conformal transformations.
This approximation is motivated
by the fact that
it is the conformal sector that determines the order of the PDE and its
pole structure.

The work is organized as follows. The background material summarizing the results on the conformally reduced flow equation obtained in \cite{Demmel:2012ub} and the heat kernel on maximally symmetric spaces are collected in sect.~\ref{sect.2} and sect.~\ref{sect.1a} respectively. In sect.~\ref{sect.4} we combine these results
into a one-parameter family of PDE's for $f(R)$-gravity, which are valid for both positive and negative background curvatures. The construction clarifies the questions related to the analytic continuation of the spherical results to negative curvature and gives a transparent meaning to the property that a solution exists for all values of $r$.
Sect.~\ref{sect.7} deals with the analytic properties of the derived flow equations.
A detailed numerical analysis of the fixed point equations is performed in sect.~\ref{sect.5}.
Our findings are summarized in sect.~\ref{sect.6}.
More technical details involving resummation and computation of the heat kernel on homogeneous spaces are discussed in app.~\ref{App.poisson} and app.~\ref{App.poly}, where a new representation of the flow equation in terms of polylogarithms is introduced.
Finally, app.~\ref{App.expand} clarifies some aspect of the computation of the relevant functional traces.

\section{The FRGE on homogeneous spaces}
\label{sect.2}
We begin by summarizing the main steps entering the derivation of
the flow equation \eqref{FRGE} for three-dimensional $f(R)$ gravity
in the conformally reduced approximation \cite{Demmel:2012ub,Demmel:2013myx}. This
setup provides an important toy model which allows
to understand the core features underlying the construction
of fixed functionals on an infinite-dimensional
truncation space, while, at the same time, being simple enough
that these features are not swamped by the technical complexity of the
complete four-dimensional analysis \cite{Machado:2007ea,Benedetti:2012dx,Dietz:2012ic,Benedetti:2013jk,Dietz:2013sba}.

The RG flow of $f(R)$-gravity considers the scale-dependence of an \emph{entire function} of coupling constants
by making the ansatz
\eq{\label{eq:ansatz}
\Gamma^{\mathrm{grav}}_k[g] = \int\mathrm{d}^3 x\sqrt{g}\; f_k(R)\, .
}
Substituting this ansatz into the FRGE and projecting the result
onto the space spanned by theories of the form $f(R)$ leads to a PDE governing the scale-dependence of $f_k(R)$. Fixed functions,
being the generalization of fixed points appearing in a finite dimensional truncation,
arise as $k$-independent solutions of this equation and are thus obtained as global solutions
a (non-linear) ODE.

The attractive feature of the $f(R)$-truncation is that the flow can be obtained by working 
with a maximally symmetric background metric $\gb_{\m\n}$. In this case the curvature of the background is completely characterized by the Ricci scalar $\Rb$
constructed from $\gb_{\m\n}$ and satisfies
\be\label{maxsym}
\Rb_{\m\n\a\b} = \frac{\Rb}{6} \left( \gb_{\m\a} \, \gb_{\n\b} - \gb_{\m\b} \, \gb_{\n\a} \right) \; , \qquad 
\Rb_{\m\n} = \frac{1}{3} \, \gb_{\m\n} \, \Rb \; , \qquad 
\bar{D}_\mu \Rb = 0 \, . 
\ee
%
In \cite{Demmel:2012ub} one particular PDE
governing the scale-dependence of $f_k(R)$ in $d=3$ was constructed
within the conformally reduced approximation \cite{Reuter:2008wj,Reuter:2008qx,Reuter:2009kq,Machado:2009ph}.
In this case
the contribution of the fluctuation fields to the r.h.s.\
of \eqref{FRGE} are restricted to the conformal mode by setting
\be\label{confapprox}
g_{\m\n} = \gb_{\m\n} + \frac{1}{d} \, \gb_{\m\n} \, \phi \, .
\ee
and considering the contribution of the fluctuation field $\phi$ only. Besides
reducing the r.h.s.\ of the FRGE to a single scalar trace this approximation
also entails that there is no need 
to supplement \eqref{eq:ansatz} by additional gauge-fixing and ghost
terms since the diffeomorphism symmetry is already broken by the conformal approximation.

Employing these approximations the construction of the operator trace entering the FRGE starts with the
computation of $\Gamma_k^{(2)}$. Substituting \eqref{confapprox} into \eqref{eq:ansatz},
the expansion of $\Gamma_k$ around the background $\gb$ takes the form
\eq{\label{eq:seriesexpansion}
\Gamma_k[\bar{g}, \phi] = \Gamma_k[\bar{g}] + \mathcal{O}(\phi) + \Gamma_k^{\mathrm{quad}}[\phi,\bar{g}]+ \mathcal{O}(\phi^3)\,.
}
This expansion \emph{should not be read} as an expansion for small values $\phi$, i.e., the fluctuations around the background
can be arbitrary large. The quadratic part of the series is \cite{Demmel:2012ub}
\eq{\label{eq:gammaquad}
\Gamma^{\rm{quad}}_k[\phi,\bar{g}] = \frac{1}{2} \, \int\mathrm{d}^3 x\sqrt{\bar{g}}\;  \phi(x) \, \Gamma^{(2)}_k[\gb] \, \phi(x)\,,
}
with kernel
\be\label{G2var}
\begin{split}
 \Gamma^{(2)}_k[\gb]
 = \frac{1}{36}&\Big[ 16 f''_k \Delta^2 + 4 \left( f'_k - 4 \bar{R} f''_k \right)\Delta +  3 \, f_k  - 4 \bar{R} f'_k  + 4 \bar{R}^2 f''_k  \Big]\,.
\end{split}
\ee
Here $\Delta \equiv - \gb^{\m\n} \bar{D}_\m \bar{D}_\n$ is the Laplacian constructed from $\gb$ and the primes denote derivatives of $f_k(\Rb)$ with respect to the background curvature scalar.
The kernel has been simplified by choosing a maximally symmetric background satisfying the identities \eqref{maxsym}.

The final ingredient entering the operator trace is the IR-cutoff $\cR_k$ which is found as follows.
Formally, we introduce the operator
\be\label{cop}
\Box \equiv - \bar{D}^2 + \Ebb
\ee
 whose eigenvalues will be used to discriminate the high-momentum and low-momentum modes. Here $\Ebb$ denotes a potential term build from $\Rb$. The
Hessian \eqref{G2var} can then be written as a function of this operator
\eq{
\Gamma^{(2)}_k = f(\Box;\dots) \,,
}
where all other dependences on $\gb$ are denoted by dots. The cutoff $\cR_k$
is then determined implicitly by demanding that
\eq{
\Gamma^{(2)}_k + \mathcal{R}_k = f(\Box + R_k;\dots) \,.
}
Thus $\mathcal{R}_k$ maintains the same tensorial structure as $\Gamma^{(2)}_k$ and dresses $\Box$ according to
\eq{
\Box \rightarrow P_k\equiv \Box + R_k \,.
}
Here, $P_k=P_k(\Box)$ plays the role of an IR-modified propagator for the scalar modes.
The function $R_k=R_k(\Box)$ is the profile function for the cutoff that contains the details of the IR-mode suppression.
We will specify the explicit form of $R_k$ later on.
Following the nomenclature of \cite{Codello:2007bd}, setting $\Ebb = 0$ corresponds to a Type Ia cutoff while $\Ebb \not = 0$ has been classified as a Type II regulator.

The flow equation for $f(R)$-gravity in $d=3$ is obtained by substituting the ansatz \eqref{eq:ansatz} into the \eqref{FRGE}
and setting $\phi = 0$ afterwards
\be\label{floww}
\int\mathrm{d}^3 x\sqrt{\gb}\; \p_t f_k(\Rb) = \frac{1}{2} \, {\rm Tr} \, W[\Box] \, . 
\ee
The function $W(z)$ reads explicitly
\eq{\label{eq:finalfloweq3sphere}
W(z)=\frac{\p_t \left(g_k\left(P_k^2 - z^2 \right) + \tilde{g}_k R_k \right)}{g_k P^2_k + \tilde{g}_k P_k +w_k},
}
with coefficients
\be\label{eq:gdef}
\begin{split}
g_k &= 16 f''_k \, , \\
\tilde{g}_k &= 4 f'_k - 16 \, \left( \bar{R} + 2 \, \Ebb \right) \, f''_k \, ,\\
w_k &= 4 \, \left(\bar{R}+2 \, \Ebb \right)^2 \, f''_k - 4 \, \left( \bar{R} + \Ebb \right) \, f'_k +3f_k\,.
\end{split}
\ee
For $\Ebb = 0$, eq.\ \eqref{eq:finalfloweq3sphere} coincides with the flow equation derived in \cite{Demmel:2012ub}. It will be
the starting point for evaluating the operator trace using the heat kernel methods reviewed in the next section.

\section{Exact heat kernels on maximally symmetric spaces}
\label{sect.1a}
A convenient tool for evaluating the operator traces
appearing on the r.h.s.\ of the FRGE is the heat kernel.
For the maximally symmetric spaces relevant for
constructing the flow equation of $f(R)$-gravity,
the exact form of the heat kernel is known and
we will summarize the relevant properties in this section.
Our exposition mainly follows \cite{Camporesi:1990wm} for spherical backgrounds while
the details of the heat kernel on $H^3$ can be found in \cite{Camporesi:1994ga}.
\subsection{The heat kernel on $S^3$}
\label{sect.1b}
In general, the heat kernel arises as the solution
of the heat equation on a manifold $\cM$
\be
(\p_s + \Delta_x) \, K(s;x, x^\prime) = 0 \, 
\ee
satisfying the boundary condition $\lim_{s \rightarrow 0} K(s;x, x^\prime) = \delta(x, x^\prime)$.
The heat kernel $K$ possesses an ``early time expansion'' for small values $s$
\be\label{earlytime}
K(s;x, x^\prime) = (4 \pi s)^{-d/2} \, \Theta(x,x^\prime) \, e^{-\sigma^2/4s} \, \sum_{n=0}^\infty \, a_n(x,x^\prime) \, s^n \, . 
\ee
Here $\sigma$ is the geodesic distance between the points $x$ and $x^\prime$ and $\Theta(x,x^\prime)$ denotes the 
van Vleck-Morette determinant. For general $\cM$ the off-diagonal heat kernel coefficients $a_n(x,x^\prime)$
can be obtained recursively \cite{Groh:2011dw}. Since the operator trace \eqref{floww} contains Laplace-type operators only,
it suffices to consider the diagonal part of \eqref{earlytime} where $K(s;x, x^\prime)$ is evaluated at the coincidence
point
\eq{\label{eq:earlytime}
K(s)\equiv K(s;x,x) \equiv \braket{x|\E^{-s\Delta}|x} = \frac{1}{\(4\pi s\)^{d/2}}\sum_n a_n(x) s^n \, . 
}
The de Witt coefficients $a_n$ can be computed by various techniques 
and we refer to \cite{Avramidi:2000bm,Groh:2011dw,Codello:2012kq,Vilkovisky:1992za} for further details and references.

In the special case where $\cM$ is the three-sphere $S^3$ the {\it normalizable} eigenmodes of the scalar Laplacian can be constructed explicitly.
Their eigenvalues $\lambda_l$ and degeneracies $D_l$ have been obtained in \cite{Rubin:1984tc}
\be\label{spectrumS3}
\lambda_l = \frac{l(l+2)R}{6} \; , \qquad D_l = (l+1)^2 \; , \qquad l = 0,1,\ldots \, .  
\ee
Moreover, the exact form of $K(s;x, x^\prime)$ can
either be obtained by direct harmonic analysis or by group theoretic considerations. In case of $S^3$ it is thereby important
that the space is compact. This entails that the heat kernel actually consists of two contributions. First,
there is the ``local heat kernel'' which reproduces the early time expansion \eqref{eq:earlytime}
evaluated on $S^3$ \cite{Avramidi:2000bm}
\eq{\label{eq:heat-kernel}
K(s)= (4\pi s)^{-3/2} \, \E^{  \frac{1}{6}R s}\, ,
}
where $R>0$ is the Ricci scalar encoding the curvature of $S^3$.
In addition to this ``local'' part there are also contributions from winding modes.
These encode the effect that the particle circles the sphere $n$ times before 
returning to its starting point after the diffusion time $s$. Combining the ``local'' and this ``topological'' contribution
yields the exact heat kernel on $S^3$ \cite{Camporesi:1990wm}
\be\label{eq:nonpert1}
K_{S^3}(s)= (4\pi s)^{-3/2} \, \E^{  \frac{1}{6}R s} \, \sum_{n = - \infty}^{\infty} \left( 1 - \tfrac{12 \pi^2 n^2}{sR} \right)  \, \E^{-\frac{6 \, n^2 \, \pi^2}{R s}} \, ,
\ee
which we give at coinciding points $x=x^\prime$.
In principle, this formula can be generalized to the off-diagonal heat kernel, but for our purpose it suffices to consider the case of coinciding points $x=x'$.
At his stage, it is illustrative to resum \eqref{eq:nonpert1}, applying the Poisson resummation formula given in app.\ \ref{App.poisson}
\be\label{eq:nonpert2}
K_{S^3}(s)= \frac{R^{3/2}}{6 \sqrt{6} \, \pi^2} \sum_{n = - \infty}^{\infty} \, n^2 \, \E^{\tfrac{1}{6} \left(1-n^2 \right) R s }\,.
\ee
The resummed heat kernel gives easy access to the asymptotic behavior of $K_{S^3}(s)$ for long diffusion times $s$,
\be\label{volume}
\lim_{s \rightarrow \infty} K_{S^3}(s) =  \frac{R^{3/2}}{12 \sqrt{6} \, \pi^2} = \left({\rm Vol}_{S^3}\right)^{-1} \, , 
\ee
where we used that the volume and curvature of $S^3$ are related by ${\rm Vol}_{S^3} = 12 \sqrt{6} \pi^2 R^{-3/2}$.
This confirms the expectation that for long diffusion time the return probability of a diffusing particle on a compact space is
uniform and
given by the inverse volume of the space. The inclusion of the ``topological'' winding modes is thereby crucial for recovering
this limit. While the analytic parts of the local heat kernel \eqref{eq:heat-kernel} and \eqref{eq:nonpert1} give rise to
the same early time expansion \eqref{eq:earlytime} their asymptotic behavior for long diffusion times are manifestly different.
This lets us expect that a well-defined flow equation for $f(R)$-gravity, valid for all values of the background curvature,
has to take the topological effect encoded in the winding modes into account.

Based on the exact heat kernel, it is straightforward to generalize eqs.\ \eqref{eq:nonpert1} or \eqref{eq:nonpert2} to the case
\eqref{cop} when the Laplace-type operator includes a covariantly constant endormorphism $\Ebb$
\be\label{mheal}
{\rm Tr} \, \E^{-s \, \Box} = \frac{1}{6 \sqrt{6} \, \pi^2} \, \int d^3x \, \sqrt{g} \, R^{3/2}  \, \E^{  \left( \frac{1}{6} \, R - \Ebb \right) \, s} \, \sum_{n = - \infty}^{\infty} \, n^2 \, \E^{- \tfrac{1}{6} \, n^2 \, R \, s } \, . 
\ee
Here $\Ebb = R/6$ is special, since for this particular choice the first exponential becomes unity and the
$s$-dependent terms are contained in the infinite sum only.
This will actually turn out to be convenient for evaluating the operator trace in \eqref{floww} in sect.\ \ref{sect.4}.

\subsection{The heat kernel on $H^3$}
\label{sect.1c}
The analogue of $S^3$ with negative scalar curvature is the hyperbolic three-space $H^3$. Similarly to $S^3$, $H^3$ is
a maximally symmetric space with $R < 0$. The crucial difference between the two spaces is that $H^3$ is non-compact. 
The spectrum of the Laplacian on $H^3$ then consists of two parts. First there is a discrete spectrum with eigenvalues $\lambda_l \le 0$.
The corresponding eigenfunctions are not normalizable, however, so that they do not give rise to a contribution to the heat kernel. The second part consists of a continuous spectrum $\rho \in [\lambda_c \, , \, \infty]$ which starts at
\be
\lambda_c = - \frac{R}{6} > 0 \, , 
\ee
and whose eigenfunctions are normalizable. Following the analysis of the spectral function \cite{Camporesi:1994ga}
for the case of $H^3$, the heat kernel encoding this spectrum is given by
\eq{\label{hkH3}
K_{H^3}(s)= (4\pi s)^{-3/2} \, \E^{  \frac{1}{6} \, R \, s} \; , \qquad R < 0 \, . 
}
For fixed curvature $R$ this expression formally reproduces the early time expansion of the heat-kernel on the sphere, while
for long diffusion times $K_{H^3}(s)$ vanishes exponentially due to the non-compactness of $H^3$. Notably, the result 
\eqref{hkH3} coincides with the analytic continuation of the \emph{local} heat kernel on $S^3$, eq.\ \eqref{eq:heat-kernel},
to negative curvature \cite{Camporesi:1990wm}. Owed to the non-compactness of $H^3$ there is no contribution
from winding modes. This feature reflects the different topologies of $H^3$ and $S^3$. In practice 
this implies that the analytic continuation of a flow equation based on the exact heat kernel \eqref{eq:nonpert2} to
negative curvature $R$ does not correctly account for the topology of $H^3$ and may thus lead to 
misleading conclusions.

Again it is straightforward to generalize \eqref{hkH3} to also include a constant endomorphism $\Ebb$. The 
heat kernel of the operator $\Box$ introduced in \eqref{cop} then reads
\be\label{hkend}
{\rm Tr} \, \E^{-s \, \Box} = (4\pi s)^{-3/2}  \int d^3x \, \sqrt{g} \,  \E^{  \left( \frac{1}{6} \, R - \Ebb \right) \, s} \, .
\ee
Similar to the spherical case, this formula simplifies considerably when setting $\Ebb = R/6$.

At this stage we make the following observation. The local parts of the heat kernel on $H^3$, eq.\ \eqref{hkH3}, and $S^3$, eq.\ \eqref{eq:heat-kernel}
are formally identical. Thus the corresponding short-time expansion agrees in both cases, implying that any finite-dimensional polynomial $f(R)$-computation 
will be insensitive to the choice of background. This feature demonstrates the ``local'' background invariance of the flow equation. 
The global topological properties of the background become visible, however, when one studies RG flows of functions like $f(R)$. In this 
case the flow equation is sensitive to the global or topological 
properties of the background. In the sequel we will show that
these topological properties are in fact essential for obtaining
solutions of the flow equation that give rise to globally defined fixed functions.  

\section{Evaluating the operator traces}
\label{sect.4}
The exact heat-kernel results reviewed in the previous section
allow the conversion of the operator equation
\eqref{floww} into a partial differential equation
encoding the scale dependence of $f_k(R)$ in a rather straightforward way.
We start by deriving the flow in the domain $R > 0$ using an $S^3$-background 
in sect.\ \ref{sect.4a} before constructing the flow valid for $R < 0$ based on
an $H^3$-background in sect.\ \ref{sect.4b}. Since the derivation is somewhat technical
the main results are summarized in tab. \ref{t.flow}.

\subsection{The flow equation on $S^3$}
\label{sect.4a}
Given the exact heat kernel \eqref{mheal}, the operator trace
in \eqref{floww} can be evaluated using Mellin-transform
techniques. These use that the trace of a general function of $\Box$ can formally related to the heat kernel
via
\eq{\label{eq:ourway}
\Tr{W(\Box)} = \int\limits_0^\infty\diff s\;\widetilde{W}(s)\Tr{ \mathrm{e}^{-s \, \Box} } \, , 
}
with
\eq{
\widetilde{W}(s) = \mathcal{L}^{-1}\left[ W \right](s)\,.
}
being the inverse Laplace transform of $W$. We then introduce the general Mellin-transform
of W as
\eq{ \label{eq:mellin}
Q_n\left[ W \right] &\equiv \int_0^\infty\diff s\;s^{-n} \, \widetilde{W}(s) \, . 
}
For $n > 0$, $Q_n[W]$ can be expressed in terms of $W$
\be
Q_n\left[ W \right] = \frac{1}{\Gamma(n)}\int_0^\infty\diff z\;z^{n-1} \, W(z)\,,
\ee
which is easily verified by expressing $W(z)$ through its Laplace transform and
using the integral representation of $\Gamma(n)$ to perform the $z$-integration.
The special case $n=0$ is obtained by comparing the definition \eqref{eq:mellin}
to the definition of the Laplace transform, yielding
\be\label{Q0}
Q_0[W] = W(0) \, . 
\ee
These formulas can be generalized by utilizing that $\mathrm{e}^{-s\alpha}$ is the Laplace representation of the translation operator on the space of functions
\eq{\label{Mellin}
Q_n \left[ W(z+\alpha) \right] = \int_0^\infty\diff s\;s^{-n} \, \widetilde{W}(s) \, \mathrm{e}^{-s\alpha}\,.
}
Here the notation means that the Mellin-transform is performed with respect to the argument $z$, instead of the full argument of $W$.

Substituting the heat kernel \eqref{mheal} into \eqref{eq:ourway} and expressing the $s$-integral in terms of 
the Mellin transform \eqref{Mellin} then yields
\spliteq{\label{wtrace}
 \operatorname{Tr} W(\Box)
 &=
 \frac{1}{12 \sqrt{6} \pi ^2} \, \int d^3x \sqrt{g}\, \sum_{n\ge 1}  \, n^2 \, R^{3/2} \,
 Q_0\left[W\!\left(z+\frac{1}{6} \left(n^2-1\right) R + \Ebb \right)\right]\\
 &=
  \sum_{n\ge 1}  \, n^2  \,
 W\!\left(\frac{1}{6} \left(n^2-1\right) R + \Ebb \right) \,.
}
Here we have used the property \eqref{Q0} and expressed the volume of $S^3$
in terms the curvature scalar via eq.\ \eqref{volume} in the second step. 

At this stage it is illuminating to study \eqref{wtrace} for $\Ebb = 0$.
Rewriting $n=l+1$ the trace becomes 
\eq{
	\Tr W\!(\Delta) =  \sum_{l\geq 0} (l+1)^2 \, W\!\( \frac{l(l+2) R}{6} \) \, = \sum_{l \geq 0} \, D_l \,  W\!\(\lambda_l\), 
}
where $\left\{ \lambda_l \right\}$  are the eigenvalues and the of the Laplace operator $\Delta$
and $D_l$ the corresponding degeneracies (see eq.\ \eqref{spectrumS3}). Thus the evaluation
of the operator trace agrees with the definition of the trace as a spectral sum \emph{over the normalizable eigenstates}. Note that the inclusion
of the ``non-local'' contribution of the winding modes in the heat kernel has been crucial for recovering
this result.

We now use the result \eqref{wtrace} to construct an explicit partial differential 
equation encoding the scale dependence of $f_k(R)$. For this purpose we first specify
the up to now general regulator $R_k$ to the optimized cutoff \cite{Litim:2001up}
\be\label{Ropt}
R_k = (k^2 - z) \, \theta(k^2-z) \, . 
\ee
Substituting \eqref{wtrace} into \eqref{floww} then becomes
\be\label{intres}
\dot{f}_k = \left. \frac{R^{3/2}}{24 \sqrt{6} \pi^2} \, \sum_{n \ge 1} n^2 \, \theta(k^2-z) \, \frac{\left( \dot{g}_k (z+k^2) + \dot{\tilde{g}}_k \right) (k^2-z) + 4 g_k k^4 + 2 \tilde{g}_k k^2}{g_k \, k^4 + \tilde{g}_k \, k^2 + w_k } \, \right|_{z = \hat{z}} \, . 
\ee
Here the dots denote a derivative with respect to $t$, the functions $g_k$, $\tilde{g}_k$ and $w_k$ have been defined in \eqref{eq:gdef} and the expression is evaluated at the eigenvalues of $\Box$
\be\label{zval}
\hat{z} \equiv  \frac{1}{6} \left(n^2-1\right) R + \Ebb \,.
\ee
The flow \eqref{intres} shows explicitly the effect of integrating out fluctuations of the background geometry.
In fact, every time the square of the RG scale $k^2$ equates a new eigenvalue $\hat{z}$
the corresponding eigenfluctuation of the background geometry is removed from the trace.
The ``integrating out'' of the fluctuations proceeds stepwise, since the spectrum of the coarse-graining operator is discrete, as manifested by the overall summation over the eigenvalues $l$.
Further, each integration enters with a relative weight factor that, modulo the multiplicity of the eigenvalues $(l+1)^2$, corresponds to the rational function on the right hand side of \eqref{intres}.

When analyzing the flow implied by \eqref{intres} it is convenient to express $R$ and $f_k(R)$ in 
terms of the dimensionless quantities
\eq{\label{dimless1}
R \equiv k^2 r\,, &\qquad 
\Ebb \equiv k^2 \ebb \,, \qquad
f_k(R)\equiv k^3 \varphi_k(R/k^2)\,.
}
The derivatives of $f_k$ and $\varphi_k$ are related by
\eq{\begin{split}\label{dimless2}\begin{aligned}
f'_k &= k \varphi'_k\,,\quad &  \dot{f}_k &=  k^3 \left(  \dot{\varphi}_k  + 3\varphi_k - 2r\varphi'_k \right)\,,\\
f''_k &= k^{-1}\varphi''_k\,,\quad  &  \dot{f}'_k &=  k \left(  \dot{\varphi}'_k  + \varphi'_k - 2r\varphi''_k \right)\,, \\
&&\dot{f}''_k &=  k^{-1} \left(  \dot{\varphi}''_k  - \varphi''_k - 2r\varphi'''_k \right)\,.
\end{aligned}\end{split}
}
In terms of these quantities \eqref{intres} becomes
\be\label{pdglS3}
\begin{split}
\dot{\varphi}_k +  3\varphi_k - 2r\varphi'_k = &  \tfrac{r^{3/2}}{24 \sqrt{6} \pi^2} \, \sum_{n \ge 1} n^2 \, \theta(1-\zeta) 
 \frac{c_1\varphi'_k +  c_2 \varphi''_k +  c_3\dot{\varphi}'_k + c_4\left(\dot{\varphi}''_k-2r\varphi'''_k\right)}{3\varphi_k+4(1-r-\ebb)\varphi'_k + 4\left(2-r-2\ebb\right)^2\varphi''_k} \,  \, , 
\end{split}
\ee
with coefficients
\be\label{ccoeff}
\begin{split}
\begin{aligned}
c_1 &= 4 \, (3 - \zeta) \, , &  
c_2 =& 16 \, (3 + \zeta^2) - 8 r \, (3 +\zeta) - 32 \ebb \, (1+\zeta) \, , \\
c_3 &= 4 \, (1-\zeta) \, , &  
c_4 =& 16 \, (1 - 2 \ebb - r + \zeta) \, (1 -  \zeta)\, . 
\end{aligned}
\end{split}
\ee
and
\be
\zeta \equiv \frac{1}{6} \left(n^2-1\right) r + \ebb \,.
\ee
Eq.\ \eqref{pdglS3} describes the RG flow of $f(R)$-gravity in the domain of \emph{positive scalar curvature} and constitutes
the main result of this subsection.

At this stage, it is illustrative to discuss the domain of validity entailed by \eqref{pdglS3}. 
When looking for fixed functionals of an RG flow, it is common folklore that the corresponding 
function $\varphi_*(r)$ should ``exist for all values of $r$''. At this point, it is important to clarify the precise meaning of this
statement. Inspecting the r.h.s.\ of eq.\ \eqref{pdglS3}, the most prominent feature is the appearance of the $\theta$-function
which leads to a step-function behavior of the equation. Every time $k^2$ crosses an eigenvalue of $\Box$ there is a new contribution. 
For concreteness, we specify the endomorphism to the two cases of interest $\ebb = 0$ and $\ebb = r/6$.  For $r \rightarrow 0$, which corresponds to $k \rightarrow \infty$ for fixed background curvature,
all eigenvalues contribute to the sum. Lowering $k$, (thereby increasing $r$) the fluctuations are integrated out successively, so that the corresponding eigenvalues no longer
contribute to the sum. In the IR $k \rightarrow 0$, $r \rightarrow \infty$ only the lowest eigenmode remains in the sum. 
Thus for $\ebb = 0$ the r.h.s.\ is non-trivial on the entire interval $r \in [0, \infty]$, while, at the same time, the constant mode is not integrated out as the flow reaches $r \rightarrow \infty$. For $\ebb = r/6$ this picture is slightly modified. In the UV, $k \rightarrow \infty$, both $r$ and $\ebb$ vanish so
that again all eigenmodes contribute to the flow. Including the endomorphism changes the IR part, however: the lowest eigenmode with $n=1$ is integrated out
at $r = 6$. For $r > 6$ all fluctuations have been integrated out and the r.h.s.\ of \eqref{pdglS3} becomes trivial. From this perspective, it makes
sense to require that the fixed function $\varphi_*(r)$ is well-defined on the $r$-interval where one actually integrates out fluctuations. This 
is the viewpoint which we will adopt in the following sections.
 
\subsection{The flow equation on $H^3$}
\label{sect.4b}
We now construct the extension of the flow equation on $S^3$ to the domain $r < 0$ using
the hyperbolic three-space $H^3$ as background manifold. Since the derivation
of \eqref{floww} holds for any maximally symmetric background, the function $W(z)$ carries over
to the negative curvature case. 

Following the strategy of the last subsection, we use the exact heat kernel on $H^3$, eq.\
\eqref{hkend}, for evaluating the operator trace. Following the steps \eqref{eq:ourway} to \eqref{Mellin}
one finds the analog of eq.\ \eqref{Mellin} valid on $H^3$
\be\label{WH3}
\operatorname{Tr} W(\Box) = \frac{1}{(4\pi)^{3/2}} \, \int d^3x \sqrt{g} \, Q_{3/2}\left[W\left(z + \hat{z}_0 \right) \right] \, . 
\ee
Here $\hat{z}_0 \equiv \hat{z}|_{n=0} = \Ebb - R/6$ denotes the restriction of \eqref{zval} to $n=0$.
Substituting the explicit form of $W(z)$ and specifying $R_k$ to the optimized cutoff,
the integrals contained in the $Q$-functionals can be reduced to the following 
basic expressions (with $m = 1,3,5$)
\be\label{eq:monint}
\int_0^\infty dz \, z^{m/2} \theta\left( k^2- z - \hat{z}_0 \right) =
\left\{ 
\begin{array}{cl}
\frac{2}{m+2} \left( k^2 - \hat{z}_0 \right)^{(m+2)/2}  & \, , \; \; k^2 - \hat{z}_0 > 0 \, , \\[1.2ex]
0 & \, , \; \; k^2 - \hat{z}_0 \le 0 \, . 
\end{array}
\right.
\ee

Substituting these integrals into \eqref{WH3} and plugging the resulting expression
into \eqref{floww} leads to the analog of eq.\ \eqref{intres} on $H^3$
\be\label{intresH3}
\dot{f}_k =  \frac{ \left( k^2-\hat{z}_0\right)^{3/2} }{210 \pi^2}  \,  \frac{70 g_k k^4 + 35 \tilde{g}_k k^2     + 2 
  \dot{g}_k  (5k^4 - 3 \hat{z}_0 k^2  -2 \hat{z}_0^2) + 7 \dot{\tilde{g}}_k (k^2 - \hat{z}_0)}{g_k \, k^4 + \tilde{g}_k \, k^2 + w_k }  \, , 
\ee
with the functions $g_k, \tilde{g}_k$ and $w_k$ defined in eq.\ \eqref{eq:gdef}.
The final result for the flow equation is obtained by writing \eqref{intresH3} in terms
of the dimensionless quantities \eqref{dimless1}
\be\label{pdglH3}
\begin{split}
\dot{\varphi}_k +  3\varphi_k - 2r\varphi'_k = &  \frac{1}{4 \pi^2} \, \left(1-\zeta_0\right)^{3/2} \,   
 \frac{\hat{c}_1\varphi'_k +  \hat{c}_2 \varphi''_k +  \hat{c}_3\dot{\varphi}'_k + \hat{c}_4\left(\dot{\varphi}''_k-2r\varphi'''_k\right)}{3\varphi_k+4(1-r-\ebb)\varphi'_k + 4\left(2-r-2\ebb\right)^2\varphi''_k} \,  \, , 
\end{split}
\ee
with coefficients
\be\label{chcoeff}
\begin{split}
\begin{aligned}
\hat{c}_1 &= \tfrac{8}{15} \left(6 - \zeta_0 \right) \, , &  
\hat{c}_2 =& -\tfrac{16}{63} \left( 49 r + 7 r \zeta_0  + 12 (\zeta_0^2 + 5 \zeta_0  -6) \right) \, , \\
\hat{c}_3 &= \tfrac{8}{15} \left(1 - \zeta_0 \right) \, , &  
\hat{c}_4 =& - \tfrac{64}{315} \left( 1 - \zeta_0 \right) \left( 14 r + 15 \zeta_0 - 15 \right) \, , 
\end{aligned}
\end{split}
\ee
and
\be
\zeta_0 \equiv \ebb - \frac{1}{6} r \,.
\ee
Eq.\ \eqref{pdglS3} describes the RG flow of $f(R)$-gravity in the domain of \emph{negative scalar curvature}.
As a consequence of \eqref{eq:monint}, the equation is only defined in the domain where $1-\zeta_0 > 0$. Outside this domain (where the square-root would become imaginary)
the r.h.s.\ is identically zero\footnote{We also confirmed this property by evaluating the functional trace via a combination of Fourier transforms and a Schr\"odinger-type regulated heat kernel.}.
For the two cases where $\ebb = 0$ and $\ebb = r/6$ this entails, that the the quantum corrections are restricted to the
intervals $r \in [-6, 0]$ and $r \in [-\infty, 0]$, respectively. In both cases the flow equation integrates out the whole spectrum of fluctuations on $H^3$,
so that both choices of the endomorphism provide a good description of the RG flow in the realm of negative scalar curvature.

%
\begin{table}[t]
\begin{tabular}{llrlc}
\hline \hline
background &  Laplacian & Lowest Eigenvalue & Domain & flow equation \\ \hline
\multirow{2}{*}{$\qquad S^3$} & $\Box = -D^2$ & $0 \qquad$  & $r \in [0, \infty]$ & \multirow{2}{*}{\eqref{pdglS3}} \\
& $\Box = -D^2 + R/6$ \quad & $R/6\qquad$ & $r \in [0, 6]$ & \\ \hline
\multirow{2}{*}{$\qquad H^3$} & $\Box = -D^2$ & $-R/6\qquad$ & $r \in [-6, 0]$ & \multirow{2}{*}{\eqref{pdglH3}} \\
& $\Box = -D^2 + R/6$ \quad & $0 \qquad$  & $r \in [-\infty, 0]$ & \\ \hline \hline
\end{tabular}
\caption{Summary of the PDEs capturing the RG flow of $f(R)$-gravity in the domains $R>0$ and $R<0$ based on two choices of the operator \eqref{cop}.
\label{t.flow}}
\end{table}

We close the section with the following remarks. The most striking difference between the flow equation for $f(R)$-gravity for $R > 0$ and $R < 0$
is that the latter case does not contain an infinite sum over winding modes. This difference originates from the different topological structure
of the compact $S^3$ and non-compact $H^3$ background. This implies that the ``analytic continuation'' of the flow equation to $R< 0$ based on 
the exact heat kernel (including winding modes) will not capture the flow on $H^3$ correctly. The corresponding equation arises from the analytic continuation 
of the flow obtained with the local heat kernel on $S^3$.
Let us furthermore stress that the r.h.s.\ 
\section{Flow equation for $f(R)$-gravity: analytical properties}
\label{sect.7}
Upon completing the derivation of the PDE encoding the
scale-dependence of $f_k(R)$, we now investigate the 
fixed functions entailed by this set of equations.
We limit our investigation to the case $\ebb = r/6$ where the flow equation integrates out
all fluctuation modes.

\subsection{Fixed function equation and its fixed singularities}
\label{sec:polestructure}
By definition, fixed functionals are given by the globally well-defined,
$k$-stationary solutions of the PDEs summarized in tab. \ref{t.flow}.
Specializing eqs.\ \eqref{pdglS3} and \eqref{pdglH3} to the case $\ebb = r/6$
and setting all $k$-derivatives to zero,
the fixed function $\varphi_*(r)$ are the globally regular solutions
of the following non-linear ODE
\be\label{odeflow}
3\varphi - 2r\varphi' = \left\{
\begin{array}{ll}
\tfrac{3 \, r^{3/2}}{4 \sqrt{6} \pi^2} \, \sum_{n \ge 1}  \theta\left(1 - \tfrac{r}{6} n^2  \right) 
 \frac{\hat{b}_1 \, n^2 + \hat{b}_2 \, n^4 + \hat{b}_3 \,n^6}{27 \varphi + 6 (6-7r)\varphi' + 16\left(3-2r\right)^2\varphi''} \, , \quad & r \in [0, 6] \, \\[2ex]
 \,  \tfrac{1}{35 \, \pi^2} \frac{252 \varphi' + 20 \left( 72 - 49 r \right) \varphi'' - 32 r  \left(15 - 14 r \right) \varphi'''}{27 \varphi + 6 (6-7r)\varphi' + 16\left(3-2r\right)^2\varphi''} \, , \quad & r \in [-\infty, 0] \, . 
\end{array}
\right.
\ee
The coefficients $\hat{b}_i$ are readily obtained from \eqref{bidefs} 
\be
\begin{split}
\hat{b}_1 = & \, 6 \varphi' + \tfrac{4}{3} \left( 18- 11 r  \right) \varphi'' 
- \tfrac{16}{3} \, r \,  \left( 3  - 4 r \right)  \varphi'''
 \, , \\
\hat{b}_2 = & \, - \tfrac{1}{9} \, r \, \left( 3 \varphi' + 10 r \varphi'' + 32 r^2 \varphi''' \right) \, , \\
\hat{b}_3 = \, & \tfrac{2}{9} \, r^2 \, \left( \varphi'' + 2 r \varphi''' \right) \, . 
\end{split} 
\ee
When analyzing the properties of the ODE, it is convenient to cast eq.\ \eqref{odeflow} into ``normal form'' by solving for the highest derivative.
This shows that the ODE is a third order equation for $\varphi (r)$.

A priori,
we thus expect that locally there is a three-parameter family of solutions.
From the theory of ODEs it is well known that such a solution does not necessarily extend to the whole domain of definition, even if the ODE is well defined everywhere.
In order to construct global solutions, it is useful to distinguish three different types of singularities that may occur in the r.h.s.\ of the ODE:
\begin{enumerate}
	\item {\bf Fixed singularities}
	\\
	Fixed singularities located at the poles $x_i$ occur whenever the r.h.s.\ of an ODE exhibits a pole structure of the form
	\eq{
		f^{(n)}(x) = \frac{ N\!( f^{(n-1)},\dots,f,x )}{(x-x_0) \dots (x-x_k)}\, .
	}
\item {\bf Moving singularities}
	\\
	Moving singularities can arise, if there is an additional differential constrained in the denominator 
	\eq{
		f^{(n)}(x) = \frac{ N\!( f^{(n-1)},\dots,f,x )}{(x-x_0) \dots (x-x_k) D( f^{(n-1)},\dots,f,x )}\, .
	}
	The r.h.s.\ becomes singular when a solution satisfies $D( f^{(n-1)},\dots,f,x ) = 0$.
\item {\bf Landau singularities}
	\\
Landau singularities cannot be immediately read off from the r.h.s.\ of the ODE: they are of a purely dynamical origin. Famous examples occurring in quantum field theory are beta functions of the type
	\eq{
		f'  \propto \frac{f^{\alpha}}{x}\, .
	}
	For $\alpha \geq 2$ there exists a finite scale $x=\Lambda_{\rm Landau}$ where $f$ diverges, even though the r.h.s.\ is smooth for all $x>0$.
\end{enumerate}
Especially fixed singularities are useful to reduce the set of solutions.
A solution is regular at a fixed singularity if, despite of the pole, the r.h.s.\ of the ODE remains finite.
This means that when the r.h.s.\ is expanded in a Laurent series at the pole $x_{\rm sing}$
\eq{\label{eq:Laurent}
	f^{(n)}(x) = \frac{ e( f^{(n-1)}(x_{\rm sing}),\dots,f(x_{\rm sing}), x_{\rm sing} )}{x-x_{\rm sing}} + \mathcal{O}\(  (x-x_{\rm sing})^0 \)\, ,
}
the principal part $e( f^{(n-1)}(x_{\rm sing}),\dots,f(x_{\rm sing}), x_{\rm sing} )=0$ has to vanish.
Technically, seeking for a regular solution translates into adding a further constraint to the boundary value problem.

\begin{figure}[t]
	\begin{center}
	\includegraphics[width=6cm]{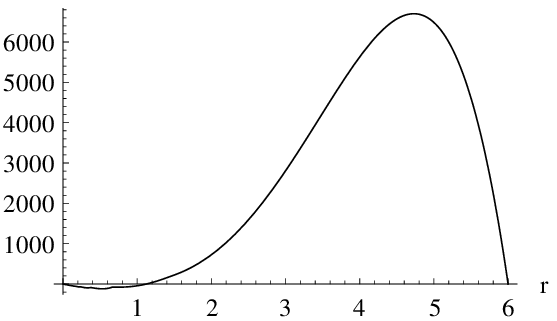}
	\includegraphics[width=6cm]{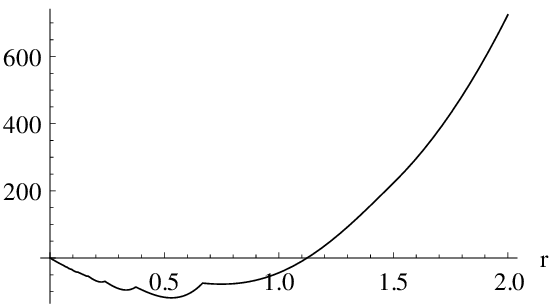}
	\caption{\label{fig:polestructure}The coefficient of $\varphi'''$ in eq.~\eqref{odeflow}. Every zero corresponds to a fixed pole, so that one finds three fixed singularities at $r_0 = 0$, $r_1 = 1.123$ and $r_2 = 6$, respectively. The kinks visible in the right diagram arise from the
stepfunction behavior of the trace when crossing an eigenvalue of $\Box$.}
\end{center}
\end{figure}

We now apply this reasoning to the pole structure of the fixed point equation \eqref{odeflow}.
The number of fixed poles can be read off as the number of zeros of the coefficient of $\varphi'''$ appearing in eq.~\eqref{odeflow}.
Technically, the zeros are determined by truncating the infinite sum to a finite sum and plotting the coefficient of $\varphi'''$ as a function of $r$, c.f.\ fig.~\ref{fig:polestructure}.
Numerically, we observe zeros at $r_0=0$, $r_2=6$ and
$r_1 \approx 1.123$.
Moreover, the flow derived on the $H^3$ background reveals that there are no poles at negative $r$. 
Hence, the number of fixed poles matches the number of initial conditions. 
The index counting \eqref{indexcounting} then suggests that the set of globally defined regular solutions is discrete.
%
%
%
%
%
%
%

Practically, the restrictions from the poles $r_{\rm sing}$ are implemented as follows. Firstly, we expand $\varphi(r)$ in a power series and substitute this series in the Laurent expansion of the flow equation \eqref{odeflow} around $r_{\rm sing}$.
This allows to determine the series coefficients $a_n$, $ n \geq 2$ in terms of the free parameters $a_0$, $a_1$
\eq{\label{eq:series-pole-0}
	\varphi(r;a_0,a_1) = a_0 + a_1 (r-r_{\rm sing}) + \sum_{n = 2}^{k} a_n (a_0,a_1) \(r-r_{\rm sing}\)^n + \mathcal{O}\(\(r-r_{\rm sing}\)^{k+1}\)\,.
}
Thus each pole reduces the dimension of the space of solutions by one.

Applying this reasoning to $r_{\rm sing}=0$ limits the possible values that the coefficients $(a_0,a_1)$ of regular solutions might take.
An explicit computation of the coefficients up to $a_{10}(a_0,a_1)$ shows that they admit a pole structure of the form
\eq{
	a_n(a_0,a_1) \propto \frac{1}{(c_1 - a_0)^{\alpha_1} \dots (c_k - a_0)^{\alpha_k}}\, ,
}
leading to singular lines in the $(a_0,a_1)$-plane.
Those lines are parallel to the $a_1$-axis and are located at
\eq{
	a_0 = \frac{2 (5-n) }{63 \pi^2}\, , \qquad n \geq 2 \,.
}
Those singular lines are densely distributed so that the complete left halfplane can be excluded from analysis.
This means that we can already exclude a negative cosmological constant.
%

%
\subsection{Smooth approximation of the spectral sum}
\label{sec:approx}
For a numerical analysis we cannot use eq.~\eqref{odeflow} as it is given there.
Therefore, we are seeking for an approximation of eq.~\eqref{odeflow} that allows us to perform a numerical integration.

\begin{figure}[t]
	\begin{center}
	\includegraphics[width=10cm]{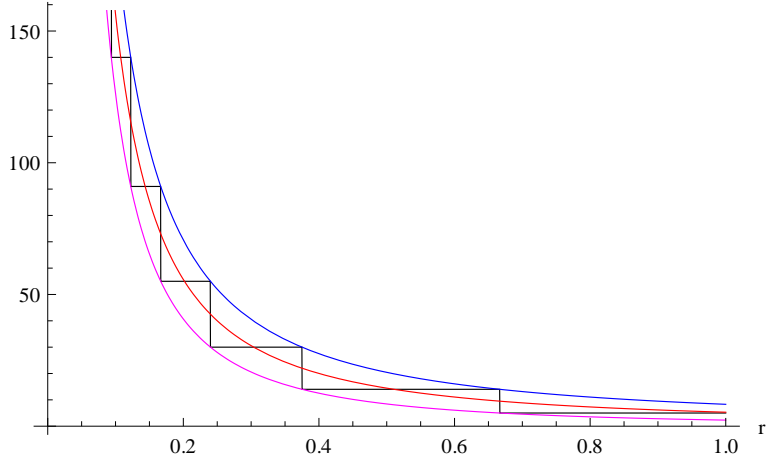}
	\caption{\label{fig:staircase-approx}
	Smooth approximations of the staircase function $\sum_n \theta\!\(1 - \frac{r}{6} n^2\) n^4$ (black line). The curves obtained
	from the replacement \eqref{stepapprox}, truncating the sum at $N_r$
and $N_r -1$ are given by the blue (top) and magenta
(bottom) curve, respectively. The average of the two approximations
resulting from eq.\ \eqref{average} gives rise to the red (middle) line.
}
\end{center}
\end{figure}

Due to the presence of the $\theta$-function, for $r>0$ the spectral sum contains only a finite number of terms and takes the form of a staircase-function, c.f.~fig.~\ref{fig:staircase-approx}.
For any term in the series to contribute, the argument of the $\theta$-function has to be greater than zero.
This gives an $r$-dependent upper boundary $N_r$ on the number of nonzero terms
\eq{\label{eq:Nofr}
	1-\frac{n^2}{6} r  > 0 \iff n< \sqrt{\frac{6}{r}} =: N_{r} \, .
}
Again, it can be seen that only for $r=0$ infinitely many terms are contributing.
Regarding the $n$-dependence, the spectral sum contains only a polynomial in $n$, as shown in eq.~\eqref{odeflow}.
Each monomial can be summed using the relation
\eq{\label{stepapprox}
	\sum_{n=1}^{M} n^k = \sum_{j=0}^{k} \binom{k}{j} \frac{B_{k-j}}{j+1} M^{j+1}\, ,\qquad B_1 = -\frac{1}{2}\,,
}
with Bernoulli numbers $B_n$.
Substituting $M\mapsto N_r = \sqrt{6 / r}$ yields a smooth approximation bounding the staircase-function from above.
A smooth approximation that bounds the sum from below is obtained by a sum from $n=1$ to $n=N_r - 1$.
Finally, we take the average of this two approximations
\eq{\label{average}
	\sum_{n=1}^{N_r} n^k \to \frac{1}{2}\( \sum_{n=1}^{N_r} n^k  + \sum_{n=1}^{N_r-1} n^k \)\,.
}
This procedure is visualized in fig.~\ref{fig:staircase-approx} for the special case $\sum_{k=1}^{N_r} n^4$.
Applying this approximation to eq.\ \eqref{odeflow} yields
\spliteq{\label{eq:flowapprox}
	3 \varphi_k - 2 r \varphi'_k =&
\left\{
\begin{array}{ll}
\frac{
	\tilde{c}_1 \varphi'_k + \tilde{c}_2 \varphi''_k  + \tilde{c}_3 \varphi'''_k
}{
	1260 \pi^2 \left(4 \left(( r-2)^2 \varphi_k'' - (r-1) \varphi'_k \right)+3 \varphi_k \right) 
}\,,  & r\in [0,6] \, ,
\\[0.4cm]
 \,  \tfrac{1}{35 \, \pi^2} \frac{252 \varphi' + 20 \left( 72 - 49 r \right) \varphi'' - 32 r  \left(15 - 14 r \right) \varphi'''}{27 \varphi + 6 (6-7r)\varphi' + 16\left(3-2r\right)^2\varphi''} \, , \quad & r \in [-\infty, 0] \, . 
\end{array}
\right.
}
with coefficients
\spliteq{
	\tilde{c}_1  &= 7 \left(r^2+15 r+144\right)\,,
	\\
	\tilde{c}_2 &= 20 \left(r^3-14 r^2-126 r+288\right) \,,
	\\
	\tilde{c}_3 &= -4 r\left(-17 r^3-35 r^2-308 r+480\right)\,.
}
The singularities, i.e. the zeros of $\tilde{c}_3$, are located at $r_{\rm sing} \in \{0,60/61,6,24\}$.
The value $r_{\rm sing} = 24$ is \emph{outside} the ODE's domain of validity and thus does not give rise to a boundary condition for a globally well-defined solution.
Thus the approximation did not change the index of the flow equation.

Another smooth approximation is obtained by keeping only the local part of the heat kernel, i.e.\ eq.~\eqref{eq:heat-kernel}, when performing the functional trace.
The result coincides with the fixed point equation on $H^3$ (second line in eq.~\eqref{odeflow}) analytically continued to $r > 0$.
We expect that this provides a good approximation for small values $r \ll 1$ and we will refer to this as the local approximation in the sequel.

\begin{figure}[t]
	\begin{center}
	\includegraphics[width=10cm]{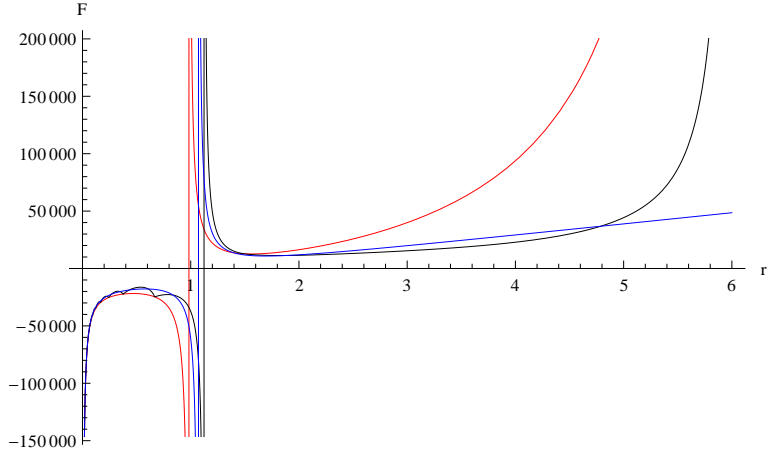}
	\caption{\label{fig:approx-quality} Numerical value of the function $F$, eq.\ \eqref{eq:approx-quality}, found by employing a finite truncation of the spectral sum in \eqref{odeflow} (black line), the smooth approximation \eqref{eq:flowapprox} (red line) and the local approximation (blue line). For $r \in (0,3/2]$ all three approximations qualitatively agree very well, while for $ r \in (3/2,6]$ both the smooth and the local approximation differ significantly from the finite truncation which is exact in this regime. }
\end{center}
\end{figure}

Before using these equations in the numerical analysis, we first estimate the quality of the approximations.
Therefore, we set $\varphi$, $\varphi'$ and $\varphi''$ on the r.h.s.\ to some fixed but arbitrary values

\eq{\label{eq:approx-quality}
	\varphi^{(3)} = F\!\( \varphi = \varphi_0, \varphi' = \varphi_1, \varphi'' = \varphi_2, r \) \equiv F(r)\, ,
}
where $\varphi_0=\varphi_1=\varphi_2={\rm const}$.
This allows to study the $r$-dependence of the r.h.s.\ of the fixed point equation.
In fig.~\ref{fig:approx-quality} we compare the resulting $F(r)$ for three different approximations of the flow equation.
The first version approximates the spectral sum in \eqref{pdglS3} by truncating the infinite sum at a finite value $N_{\rm max}$ (black line). Note that according to \eqref{eq:Nofr} there exists a $r_{\rm min}$ such that the resulting flow equation is exact for $r > r_{\rm min}$.
For this reason it is regarded as a reference for the other two approximations in this domain.
This result is compared to the smooth approximation \eqref{eq:flowapprox} (red line) and the local approximation \eqref{pdglr2} (blue line).
It can be seen that for $r \in (0,3/2]$ all three approximations qualitatively agree very well, while for $ r \in (3/2,6]$ both the smooth and the local approximation differ significantly.
The smooth approximation is capable of reproducing the pole at $r=6$, while the local approximation misses this crucial feature and is thus not 
suitable for describing the large $r$ behavior. 
Choosing different values for the constants $\varphi_0$, $\varphi_1$ and $\varphi_2$ leads to very different diagrams, but the qualitative features remain the same.
Therefore, we conclude that the observations are independent of the choice of the constants.
This analysis suggests to use a piecewise definition of the fixed point equation for the numerical analysis.
On $(0,3/2]$ the smooth approximation can safely be used, while on $(3/2,6]$ one should work with the exact equation.
This is the strategy that will be employed in the numerical analysis of the next section.
%
%
%
\section{Flow equation for $f(R)$-gravity: numerical analysis}
\label{sect.5}
%
%
We continue our analysis of the ODE \eqref{odeflow}
by constructing explicit solutions with the help of numerical techniques.
In this way we are able to identify two globally well-defined fixed functions
whose deformations are studied in Sect.\ \ref{sect:6.2}.
%
%
\subsection{Numerical construction of fixed functions}
%
We start with a numerical search for globally well-defined
solutions of the fixed function equation \eqref{odeflow}.
Since all fixed singularities are located at $r \ge 0$,
we first investigate the positive curvature domain and subsequently
extend potential fixed functions to the domain of negative curvature.

In order to make eq.\ \eqref{odeflow} amenable to a numerical treatment,
we implement the following approximations. First, we split the domain 
of definition $r \in [0,6]$ into a ``UV-interval'' $[0,3/2)$
and an ``IR-interval'' $[3/2,6]$. On the UV-interval we then
employ the smooth approximation eq. \eqref{eq:flowapprox} 
which captures all essential features of the infinite sum.
On the IR-interval the spectral sum reduces to the term with $n=1$ and we use the exact equation \eqref{odeflow}.
Implementing the IR-asymptotics correctly thereby turns out to be crucial for the existence 
of fixed functions.
 
As was pointed out above, the fixed function equation is of third order.
Locally, it thus gives rise to a three-parameter family of solutions.
Globally, this set of solutions is restricted by boundary conditions imposed
at the three fixed singularities $r_{i,{\rm sing}} = \{ \, 0 \, , \, 60/61 \, , \, 6 \, \}$.
The precise form of these boundary conditions is obtained by
expanding the fixed function equation in a Laurent series at the singularities
and demanding that the principle part of the series, containing the residues $e_i$, vanishes.
For the three fixed singularities this implies the vanishing of the combinations
\eq{\label{eq:cond1}
	e_0 & =  105 \pi ^2 \varphi \left(16 \varphi''+ 4 \varphi' +3 \varphi\right)-160 \varphi''+28 \varphi' \,,
	\\\label{eq:cond2}
	e_1 &=  45 \pi ^2 \left(61 \varphi-40 \varphi'\right) \left(2352 \varphi''-732 \varphi'+3721 \varphi\right)
		-2015880 \varphi''-939949 \varphi' \,,
		\\\label{eq:cond3}
	e_2 &=	-9 \pi ^2 \left(\varphi-4 \varphi'\right) \left(48 \varphi''-8 \varphi'+\varphi\right) - 48 \varphi''+ 2 \varphi' \, ,
}
evaluated at $r = r_{i,{\rm sing}}$. Here the residue $e_2$ at $r_{2,{\rm sing}}$ is obtained from the exact equation \eqref{odeflow}, while the residues at $r_{0,{\rm sing}}$ and $r_{1, {\rm sing}}$ are extracted from the smooth approximation eq.~\eqref{eq:flowapprox}.

The construction of the numerical solutions starts at the first order pole at $r_{0, {\rm sing}} = 0$ which corresponds 
to the deep UV. We impose that $\varphi(r)$ has an analytic expansion at this point
\be\label{vpana}
\varphi(r) = \sum_{n=0}^\infty \, a_n \, r^n \, ,
\ee
with hitherto undetermined coefficients $a_n$. Substituting this series into \eqref{eq:cond1} fixes $a_2$ as a function of $a_0, a_1$. The coefficients
$a_n$, $n \ge 3$ are obtained recursively from the equations read off from the positive powers of $r$ in the expansion of \eqref{eq:flowapprox}. Thus
the vanishing of the residue $e_0$ reduces the number of regular solutions to a two-parameter family $\varphi(r)$ whose initial conditions
are conveniently parameterized by $a_0,a_1$.

In terms of constructing numerical solutions, it is impossible to impose initial conditions directly at a fixed singularity. In order to
bypass this obstacle, we use the analytic expansion \eqref{vpana} to map the initial conditions encoded in the pair $(a_0, a_1)$ to initial data
at $r = \varepsilon$, 
\be\label{initdata}
\varphi_{\rm init}(\varepsilon) = \varphi (\varepsilon;a_0,a_1)\,, \quad 
\varphi'_{\rm init}(\varepsilon) = \varphi' (\varepsilon;a_0,a_1)\,, \quad
\varphi''_{\rm init}(\varepsilon) = \varphi'' (\varepsilon;a_0,a_1)\,,
\ee
where the r.h.s.\ is obtained by evaluating the analytic expansion to an sufficiently high order.
For practical purposes we chose $\varepsilon = 10^{-4}$ and checked that the results are stable
with respect to changing the order of the expansion and as long as $\varepsilon$ is not too small.

\begin{figure}[t]
	\begin{center}
	\includegraphics[width=10cm]{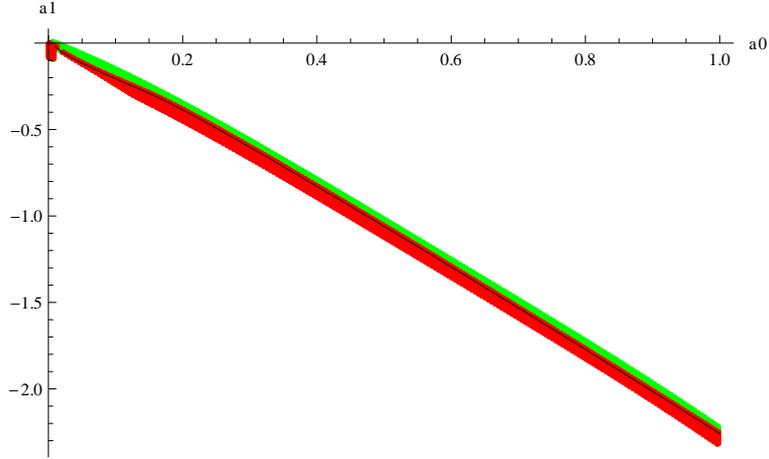}
	\caption{\label{fig:a0a1plane-colour} The $(a_0,a_1)$-plane of solutions regular at $r=0$. 
	Every point of the green area corresponds to a pair $(a_0,a_1)$, for which $e_1(a_0, a_1)$ is positive while red points correspond to solutions, for which $e_1(a_0, a_1)$ is negative.
The black line parametrizes those pairs $(a_0,a_1)$, for which \eqref{eq:cond2} vanishes (within numerical precision).
Accordingly, the black line corresponds to solutions which are regular at $r_{0, {\rm sing}}$ and $r_{1, {\rm sing}}$.}
\end{center}
\end{figure}

The initial data \eqref{initdata} serves as input for a numerical shooting method which
extends the two-parameter family of solutions to the interval $[0, r_{1, {\rm sing}}-\varepsilon_1]$.
Notably, for most values $(a_0, a_1)$ the solutions do not extend up to the second fixed singularity, but
terminate in a moving singularity at $r_{\rm term} < r_{1, {\rm sing}}$. The solutions
that reach $r_{1, {\rm sing}}-\varepsilon_1$ are matched to an analytic expansion of $\varphi(r)$ at $r_{1, {\rm sing}}$, 
from which we obtain numerical values for $\varphi(r_{1, {\rm sing}})$, $\varphi'(r_{1, {\rm sing}})$, and $\varphi''(r_{1, {\rm sing}})$,
as functions of $(a_0, a_1)$. Inserting these values into the condition \eqref{eq:cond2} then defines an implicit function
$e_1(a_0, a_1)$. Regular solutions fulfill $e_1(a_0,a_1)=0$.
The shooting method now varies $a_0$ and $a_1$ and seeks for values $(a_0, a_1)$ which satisfy this condition.
The result is shown in fig. \ref{fig:a0a1plane-colour}.
The result illustrates that the regular solutions passing $r_{1, {\rm sing}}$ 
constitutes a one dimensional manifold (black line), which we refer to as regular line.
This line can conveniently be parameterized by one single parameter which we take to be $a_0$.
Note that the appearance of the line, fixing one of the free parameters, is in complete 
agreement with the singularity counting theorem advocated in the introduction.

The next step repeats the procedure above and applies the shooting
method with the goal of extending the one-parameter family of regular
solutions shown in fig.\ \ref{fig:a0a1plane-colour} to the entire 
domain of definition $[0, 6]$. The initial conditions for the second numerical
integration are obtained by evaluating the analytic expansion of $\varphi(r)$
at $r_{1, {\rm sing}}$ to leading order in $\varepsilon$
\be
\begin{split}
\varphi_{\rm init}(r_{1, {\rm sing}} + \varepsilon) = & \varphi(r_{1, {\rm sing}}-\varepsilon; a_0) \, , \qquad
\varphi_{\rm init}'(r_{1, {\rm sing}} + \varepsilon) =  \varphi'(r_{1, {\rm sing}}-\varepsilon; a_0) \, ,  \\
\varphi_{\rm init}''(r_{1, {\rm sing}} + \varepsilon) = & \varphi''(r_{1, {\rm sing}}-\varepsilon; a_0) \, . 
\end{split}
\ee
\begin{figure}[t]
\begin{center}
	\includegraphics[width=7cm]{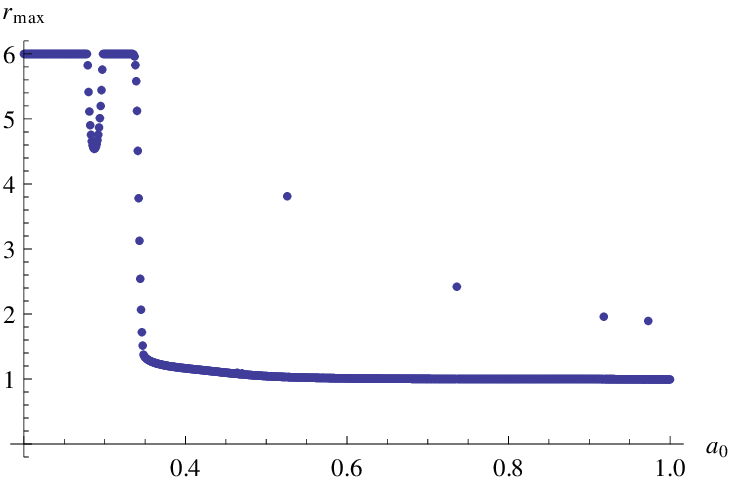}
	\includegraphics[width=7cm]{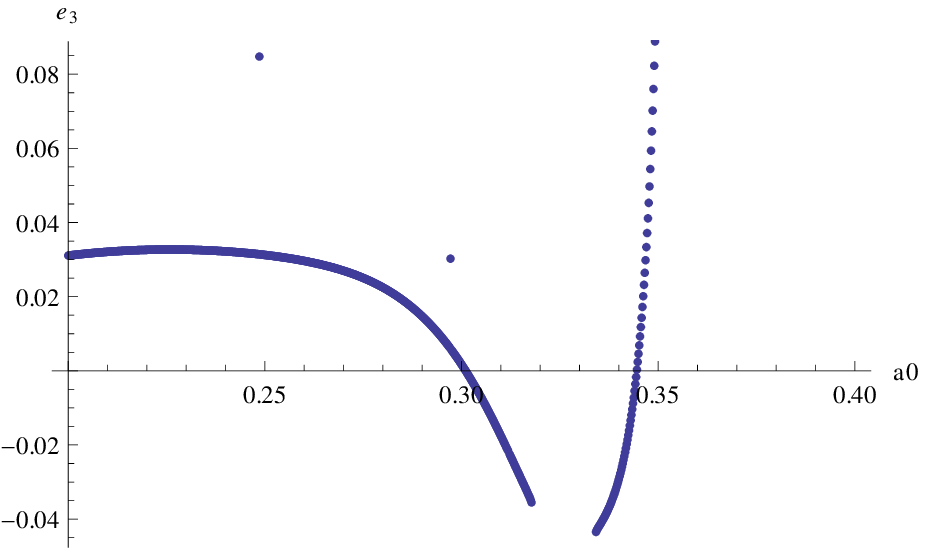}
	\caption{\label{fig:error3}
	The left panel shows the maximal interval of existence $[0,r_{\rm max})$ of the regular solutions $\varphi(r;a_0,a_1)$ as a function of $a_0$. The regularity condition $e_3(a_0)$ given in eq.\
\eqref{eq:cond3} is shown in the right panel. The latter displays two zeros corresponding to two isolated fixed functions.}
\end{center}
\end{figure}
Again we numerically integrate these initial conditions up to $r_{2, {\rm sing}}-\varepsilon$. 
For large values $a_0$ the solutions do not extend to the third pole and terminate in moving singularities at $r_{\rm max}$.
This feature is illustrated in the left part of fig.~\ref{fig:error3} which shows $r_{\rm max}$ as a function of $a_0$.
For small values $a_0$ one, however, obtains numerical solutions which do extend to $r_{2, {\rm sing}}-\varepsilon$.
Inserting the value of these functions and its derivatives obtained through the numerical integration into
eq.~\eqref{eq:cond3} again defines an implicit function $e_3(a_0)$. In order to be completely regular
on the domain $[0,6]$, these solutions then have to satisfy  $e_3(a_0) = 0$. This function is shown
in the right panel of fig.\ \ref{fig:error3}. This figure establishes that there are {\it two} distinguished 
values for $a_0$ where the last regularity condition is satisfied. Thus we succeeded to construct
 two distinct  fixed functions which are regular on the domain $[0,6]$. We will denote these solutions by
$\varphi_{1}$ and $\varphi_2$. The corresponding values of the points $(a_0,a_1)$ are listed in tab.~\ref{tab:num-values}.

\begin{figure}[t]
\begin{center}
	\includegraphics[width=10cm]{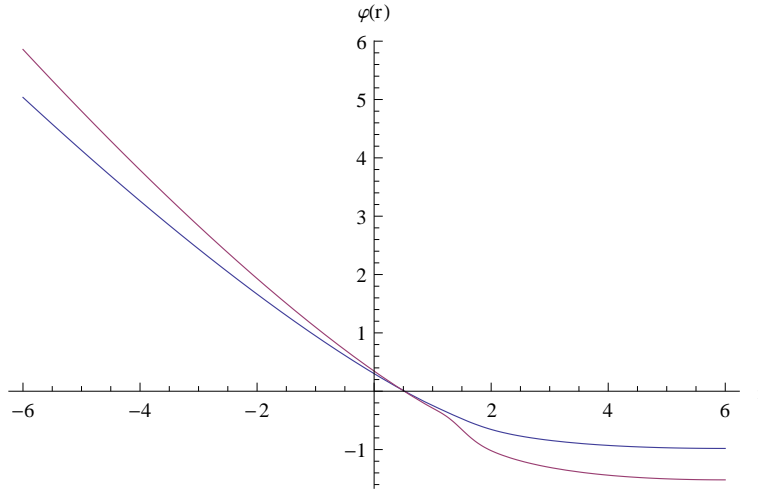}
	\caption{\label{fig:NumSolF1F2} The full numerical solutions $\varphi_{1}(r)$ (blue line) and $\varphi_{2}(r)$ (magenta line). Both solutions
	are regular on the entire domain $[-\infty, 6]$ and satisfy all properties required from genuine fixed functions.
}
\end{center}
\end{figure}

In the final step, we extend the numerical solution also to negative values of $r$ using the exact
equation \eqref{odeflow} in this domain. Since there are no fixed poles for $r<0$ we can
safely use eq.~\eqref{odeflow} to extend to $\varphi_{1}$ and $\varphi_2$ to the domain $[-\infty, 0]$.
The full numerical solutions are displayed in fig. \ref{fig:NumSolF1F2}.
Quite remarkably, {\it both solutions} have a regular extension on the entire domain of negative values $r$. 
Thus they satisfy all the requirements imposed on genuine fixed functions for $f(r)$-gravity in three dimensions.
This constitutes the main result of this section.

As pointed out before, when expanding around $r=0$, all series coefficients can recursively be expressed as functions of $a_0$ and $a_1$, c.f.\ eq.~\eqref{eq:series-pole-0}.
Having $(a_0,a_1)$ of the two fixed functions allows for computing a series expansion of $\varphi_1$ and $\varphi_2$ to any desired order.
Therefore, it is possible to compute every coupling constant arising in an polynomial expansion in $r$ around $r=0$. For comparison with
earlier computations, it is illustrative to relate the first three expansion coefficients to the (dimensionless) Newton's constant $g$, the cosmological constant $\lambda$ and the 
inverse of the $R^2$ coupling $b$, which typically parametrize the set of coupling constants in the $R^2$ truncation, by
\be
g^* = - \left(16 \pi a_1^*\right)^{-1} \, , \qquad \lambda^* = - a_0^*/(2 a_1^*) \, , \qquad b^* = (a_2^*)^{-1} \, . 
\ee
The numerical values of these couplings are displayed in tab.~\ref{tab:num-values}.
Notably, both fixed point solutions give rise to a positive Newton's constant and cosmological
constant in agreement with the expectations from studying finite-dimensional RG flows. Since $g^*$ and $\lambda^*$ by themselves display a rather strong dependence on the
unphysical regulator, we also give the value of the 
 universal scaling variable  \cite{Lauscher:2001ya}
\eq{
	\tau^* = \lambda^* \left( g^* \right)^2 \, . 
}
The values obtained for $\tau^*$ and $b^*$ are actually similar to the ones 
obtained in a $R^2$ truncation \cite{Rechenberger:2012pm}, where $\tau^* =0.00067$ was found for the physical fixed point. This
builds up further confidence in the validity of our findings, since  $\varphi_{1}(r)$ and  $\varphi_{2}(r)$ share 
many of the characteristic features of the NGFPs encountered in previous computations.
\begin{table}
\centering
\begin{tabular}{ccccccc}
\hline \hline
Fixed function & $a_0^*$ & $a_1^*$ & $\lambda^*$ &$g^*$ & $b^*$ & $\tau^*$ \\  \hline
$\varphi_{1}$  & 0.3011 & -0.6041 &0.2492&0.0329&157.18&0.000270 \\
$\varphi_{2}$  & 0.3449  & -0.7026 &0.2454&0.0283&131.22& 0.000197\\
\hline \hline
\end{tabular}
\caption{\label{tab:num-values} The numerical values and characteristic features of the two distinct fixed functions.}
\end{table}

%
\subsection{Identifying the relevant deformations}
\label{sect:6.2}
%
%
In the previous section we constructed two distinct regular fixed functions $\varphi_1$ and $\varphi_2$. In order to
further strengthen the connection of these solutions with the ones obtained from finite-dimensional truncations,
 we proceed by determining the number of relevant directions in the UV, i.e.\ for $k\to\infty$ or $r\to 0$.
For this purpose we expand $\varphi_k$ and $\dot{\varphi}_k$ around $r=0$
\spliteq{\label{expand}
	\varphi_k (r) &= \sum_{i \geq 0}^N\; g_i\, r^i\,, \qquad
	\dot{\varphi}_k(r) = \sum_{i \geq 0}^N\; \beta_i\, r^i\,
}
where the $g_i$ are the $k$-dependent coupling constants appearing in the expansion and the $\beta_i$ denote the corresponding beta functions.
This expansion is then substituted into the  local approximation of the full flow equation obtained from restricting \eqref{pdglr} to the first line
and specializing to $\ebb = r/6$.
Expanding this equation up to a fixed order in $r$ yields a coupled system of equations that can be solved for the beta functions.
The fixed points then appear as algebraic solutions of this system of equations satisfying $\beta_i(g_i^*) = 0$. 
Contrary to the usual polynomial expansion where the system is closed by setting the highest order coefficients $g_{N+1} = g_{N+2}=0$, we specifically dial to the fixed functions $\varphi_1$ and $\varphi_2$ by fixing the lowest coefficients $(a_0^*, a_1^*)$ to be the ones given in tab.\ \ref{tab:num-values} and
subsequently determining the higher order coefficients $a_n^*$ by solving
the recursion relations arising from the local approximation of the fixed
function equation.

The information on the relevant deformations of the solution is encoded in the
 stability matrix, defined as the Jacobian matrix of the beta functions
\eq{
	(\mathbf{M})_{ij}= \left. \tfrac{\del \beta_i}{\del g_j} \right|_{g_i = g_i^*}\,,
}
evaluated at the fixed point values. The stability coefficients $\theta_i$ are defined as minus the eigenvalues of $\mathbf{M}$,
so that a $\theta_i$ with a positive real part denotes a relevant deformation. 
\begin{table}
\centering
\begin{tabular}{cccccccc}
\hline \hline
 & $\theta_1$ & $\theta_2$ & $\theta_3$ &$\theta_4$ & $\theta_5$ & $\theta_6$ &$\theta_7$\\  \hline
$N=1$  & $2.95$ & $0.96$ &&&& \\
$N=3$  & \multicolumn{2}{c}{$1.64  \pm 0.48\, \i$}  & $-0.12$ &$-2.12$&&&\\
$N=5$  & \multicolumn{2}{c}{$2.45 \pm 2.47\, \i$} &$-0.29$&$-2.52$& $-5.41$&$-8.56$ \\
\hline
$N=2$ & $2.85$ & \multicolumn{2}{c}{$0.29\pm 0.66\, \i $} \\
$N=4$  & $2.79$ & \multicolumn{2}{c}{$0.69 \pm 2.34\, \i$} & $-2.13$ &$ -6.70$ &  \\
$N=6$  & $3.74$ & \multicolumn{2}{c}{ $0.31 \pm 4.22\,\i$} &$ -1.35$ & $-5.35$ & \multicolumn{2}{c}{$-9.32 \pm 5.59\,\i$}\\
\hline \hline
\end{tabular}
\caption{\label{tab:thetas1} Stability coefficients $\theta$ obtained from the expansion \eqref{expand} for $\varphi_1$ up to order $N$.}
\end{table}
\begin{table}
\centering
\begin{tabular}{cccccccc}
\hline \hline
  & $\theta_1$ & $\theta_2$ & $\theta_3$ &$\theta_4$ & $\theta_5$ & $\theta_6$ &$\theta_7$\\  \hline \hline
$N=1$  & $2.96$ &$0.97$  \\
$N=3$  & \multicolumn{2}{c}{$1.74\pm 0.63\,\i$} & $0.11$ & $-1.80$ \\
$N=5$  &  \multicolumn{2}{c}{$2.88\pm 2.61\,\i$} & $-0.32$ & $-1.69$ & $-4.96$& $-7.68$\\
\hline
$N=2$  &  $2.87$ & \multicolumn{2}{c}{$0.38\pm 0.72\,\i$}\\
$N=4$  & $2.86$ & \multicolumn{2}{c}{$ 1.09\pm 2.44\, \i $} & $-1.67$ & $-6.33$  \\
$N=6$  &  $3.83$ & \multicolumn{2}{c}{$1.30\pm 4.20\,\i$} & $-0.80$ & $-4.62$ & \multicolumn{2}{c}{$-8.93\pm 6.03\, \i$} \\
\hline \hline
\end{tabular}
\caption{\label{tab:thetas2} Stability coefficients $\theta$ obtained from the expansion \eqref{expand} for $\varphi_2$ up to order $N$.}
\end{table}

The information on the stability coefficients obtained this way is summarized in 
tab.~\ref{tab:thetas1} and tab.~\ref{tab:thetas2} for $\varphi_1$ and $\varphi_2$, respectively.
Notably, determining the $\theta_i$ for $N \ge 6$ becomes numerically challenging and, in particular, sensitive to the numerical accuracy to which $a_0^*$, $a_1^*$ are known.
We find that the stability coefficients seem to organize themselves into two subseries,
depending on whether the expansion \eqref{expand} terminates with $N$ even or odd. The
data supports two or three relevant deformations, depending on the particular value of $N$.
The convergence of the stability coefficients is much less obvious than in the corresponding
polynomial expansions of $f(R)$-gravity in four dimensions \cite{Codello:2007bd,Machado:2007ea,Codello:2008vh,Bonanno:2010bt,Falls:2013bv}. It would be very interesting
to see if this feature can be traced back to the omission of the transverse traceless fluctuation modes
or is linked to working in an odd-dimensional spacetime. For the time being we, however, conclude with
the observation that the stability analysis of the fixed functions $\varphi_1$ and $\varphi_2$
gives rise to a  finite dimensional critical surface, supporting the conjecture \cite{Benedetti:2013jk}.
\section{Summary and conclusions}
\label{sect.6}
In this paper we used the functional renormalization group equation for the effective average action $\Gamma_k$ \cite{Reuter:1996cp} to investigate the RG flow of $f(R)$-gravity in three-dimensional, conformally reduced Quantum Einstein Gravity (QEG). This setup provides an important toy model which exhibits many features encountered when studying the RG flow of $f(R)$-theories in fledged QEG in four dimensions without being swamped by the overwhelming technical complexity of the later. Our simplified setting may therefore provide important guidance for identifying relevant structures that need to be taken into account when studying the fixed point structure of gravity in infinite-dimensional truncation spaces.

At this stage, it is worthwhile to summarize the picture that has been put together in the series of studies \cite{Machado:2007ea,Benedetti:2012dx,Demmel:2012ub,Benedetti:2013jk,Dietz:2012ic,Demmel:2013myx,Benedetti:2013nya,Dietz:2013sba,Bridle:2013sra} focusing on the RG flow of $f(R)$-gravity. The scale-dependence of $f_k(R)$ obtained from projecting the flow equation for the effective average action on gravitational actions of $f(R)$-type is a complicated non-linear partial differential equation (PDE), which is of first order in the $k$-derivative and third order in the derivatives with respect to the dimensionless curvature scalar $r \equiv R/k^2$. This setting is more complex than the scalar case where a similar ``local polynomial approximation'' results in a second order PDE for the scalar potential \cite{Wetterich:1992yh,Morris:1994ie,Morris:1998da}.
In this setting fixed functions are, by definition, globally well defined $k$-independent solutions of this PDE. Thus they are governed by a non-linear third order ordinary differential equation (ODE). Naively, this setting allows for a three-dimensional set of solutions, characterized by the initial conditions of the ODE. The requirement that the solution should be globally well-defined then poses constraints on the admissible parameters. Typically the domain of definition of the ODE contains fixed singularities. Demanding regularity at these points fixes the free parameters. In \cite{Dietz:2012ic} it has then been argued that 
\be
\mbox{dimension of solution space} = \mbox{order of the ODE} - \mbox{total order of singularities}\,.
\ee
Balancing the order to the ODE with these boundary conditions, one expects that fixed functionals appear as isolated points in the space of all solutions.

At this stage it is necessary to give a precise meaning to the statement that the fixed function should be globally well defined. Naively, one would expect that this entails that the solution should be regular on the whole interval $r \in [-\infty, \infty]$. This turns out to be too stringent, however. Thus we advocate that it is sufficient that the solution exists on the interval in which the quantum fluctuations are integrated out. The precise domain of definition depends on the choice of cutoff operators, and typically contains a subspace of the real line only (cf.\ tab.\ \ref{t.flow}).

Owed to the complex nature of the ODE solutions have to be constructed numerically in a ``bottom-up'' way. This construction starts at $r = 0$ (corresponding to $k \rightarrow \infty$ for fixed background curvature $R$) and imposes that the solution has a polynomial expansion.\footnote{In terms of fixing parameters through regularity conditions, it may be preferential to start in the IR at large values $r$. In this case it is not clear that the solution should admit an analytic expansion and one typically encounters non-analytic terms \cite{Dietz:2012ic,Demmel:2013myx}.}
Subsequently, the solution is extended towards the IR, fixing the free parameters when passing through a singular locus. While, so far, it has not been possible to construct a fixed functional satisfying these criteria in full-fledged QEG in four dimensions, it was argued in \cite{Benedetti:2013jk} that, given that such a functional exists, it will automatically come with a finite number of relevant deformations, therefore ensuring the predictivity of the construction.

One of the major contributions towards establishing the overall picture underlying the construction of fixed functionals provided by the present work is the derivation of the flow equation of $f(R)$ gravity on maximally symmetric spaces with positive (compact three-sphere) and negative scalar curvature. Utilizing a compact spherical background $S^3$ and non-compact hyperbolic three-space $H^3$ we obtained the set of PDEs summarized in tab.\ \ref{t.flow} which cover the scale-dependence of $f(R)$-gravity on the $r$-interval including all fluctuation modes. This derivation revealed that the flow equation is strongly modified by the topology of the background. The correct PDE in the domain $r < 0$ can be obtained by analytic continuation of the spherical result.
This extension has to be performed with care, however, using the result for the ``local heat kernel'' on $S^3$ only. This feature explicitly demonstrates the topology dependence of the gravitational RG flow. The use of the local heat kernel (usually in the early time expansion), ensures that finite-dimensional truncations exhibit a ``local background covariance''. These computations do not feel the global properties of the background and the beta functions are universal in the sense that they do not depend on the chosen background metric \cite{Benedetti:2010nr}. This feature is lost when performing computations at the level of functions, where global properties of the heat kernel play a crucial role for producing the correct asymptotics of the flow equation.\footnote{We expect that a similar ``topological background dependence'' also arises in the context of the transverse traceless decomposition of the gravitational fluctuations \cite{Lauscher:2001ya}. In this case the use of background geometries possessing killing vectors and conformal killing vectors induce additional terms in the flow equation, related to zero modes of the decomposition.}
As an amazing byproduct \eqref{intres} illustrates the working of the FRG on a background three-sphere. In this case, fluctuations and multiplicity are known and one can explicitly trace how the FRG realizes the ``integrating out'' of the discrete levels of fluctuation modes.

The numerical analysis of the non-linear ODE \eqref{odeflow} encoding the fixed functions on the space of $f(R)$-truncations is technically rather demanding. By combining analytical and numerical techniques, this analysis established the existence of two isolated and globally defined fixed functions based on the general guiding principles established above. The two solutions give rise to a positive dimensionless Newton's constant $g_*$ and cosmological constant $\lambda_*$. Even though the analysis has been carried out with the conformally reduced approximation, the properties of these solutions, including the universal product $\lambda_* g_*^2$, the coupling constant of the $R^2$ coefficient and critical exponents turn out to be similar to the results obtained within full QEG in a three-dimensional spacetime \cite{Rechenberger:2012pm}. Based on this matching we are confident that the fixed points observed in finite-dimensional polynomial expansions of an $f(R)$-type ansatz for $\Gamma_k$ \cite{Codello:2007bd,Bonanno:2010bt,Machado:2007ea,Codello:2008vh,Falls:2013bv} can be extended to the realm of fixed functions. We expect that the picture developed above will provide important guidance for investigating the existence and properties of fixed functionals in realistic models including the $f(R)$-approximation of full QEG in three and four-dimensions. We hope to come back to this point in future works.

\acknowledgments
We thank D.\ Benedetti, A.\ Bonanno, T.\ Morris and M.\ Reuter for helpful discussions. The research of F.~S.\ and O.~Z.\ is supported by the Deutsche Forschungsgemeinschaft (DFG)
within the Emmy-Noether program (Grant SA/1975 1-1).

\begin{appendix}
\section{Poisson resummation}
\label{App.poisson}
In order for evaluating the functional traces in sect. \ref{sect.2}, we made use of the Poisson resummation formula. The basic 
ingredient for this resummation is the identity for the Dirac delta-distribution
\be
\sum_{n \in \mathbb{Z}} \delta(y - n a) = \frac{1}{a} \, \sum_{n \in \mathbb{Z}} \, e^{2\pi \i ny/a} \, , \; a \in \mathbb{R}^+ \, . 
\ee
Multiplying with an arbitrary function $f(x+y)$ and integrating over $y \in \mathbb{R}$ gives the Poisson resummation formula
\be\label{prf}
\sum_{n \in \mathbb{Z}} \, f(x+na) = \frac{1}{a} \, \sum_{n \in \mathbb{Z}} \, \tilde{f}(2\pi n/a) \, e^{2\pi \i nx/a}
\ee
Here $f(x)$ and $\tilde{f}(k)$ are related by the Fourier transform
\be
\tilde{f}(k) = \int_{-\infty}^\infty \, \diff x \, f(x) \, e^{-ikx} \, , \qquad f(x) = \frac{1}{2\pi} \, \int_{-\infty}^\infty \diff k \, \tilde{f}(k) \, e^{ikx} \, .   
\ee
%
\section{The flow equation on $S^3$ in terms of polylogarithms}
\label{App.poly}
While providing an intuitive picture in terms of integrating out fluctuations 
and providing a good starting point for the numerical search for fixed functions,
the connection of the flow equation \eqref{pdglS3} to the local and non-local 
parts of the heat kernel is not visible. In this section we clarify this relation by undoing the Poisson resummation
which allowed to derive \eqref{pdglS3}. Arguably, it is this equation that should be used in the polynomial expansion
of the flow equation at $r=0$, since in this case, the relation to the early-time expansion of the heat kernel
becomes manifest.

We start by recasting the PDE \eqref{pdglS3} in a form where the $n$-dependence becomes manifest
\be\label{pdglr1}
\begin{split}
\dot{\varphi}_k & +  3\varphi_k - 2r\varphi'_k = \\ &   \tfrac{r^{3/2}}{24 \sqrt{6} \pi^2} \, \sum_{n = - \infty}^\infty  \, \theta\left(\tilde{\zeta}-\tfrac{r}{6} n^2 \right) 
 \frac{b_1 \, n^2 + b_2 \, n^4 + b_3 \, n^6}{3\varphi_k+4(1-r-\ebb)\varphi'_k + 4\left(2-r-2\ebb\right)^2\varphi''_k} \,  \, . 
\end{split}
\ee
The sum now runs over all values $n \in \mathbb Z$, $\tilde{\zeta} \equiv 1+ \tfrac{1}{6}r-\ebb$, and the coefficients $b_i$ are obtained from a series expansion of the numerator appearing in \eqref{pdglS3}
\be\label{bidefs}
\begin{split}
b_1 = & \, \big(4 + 2\tilde{\zeta} \big) \varphi' + 2 \tilde{\zeta} \dot{\varphi}' + \tfrac{8}{3} \tilde{\zeta} \big( 3 \tilde{\zeta} - 4 r \big) \big(\dot{\varphi}'' - 2 r \varphi'''\big)
- \tfrac{4}{3} \big( 6 \tilde{\zeta}^2 - 5 r \tilde{\zeta} + 8 (2 r - 3 \tilde{\zeta}) \big) \varphi'' \, , \\
b_2 = & \, - \tfrac{1}{9} \, r \, \left( 3 \dot{\varphi}'_k - 16 r \dot{\varphi}''_k + 3 \varphi' + 10 r \varphi'' + 32 r^2 \varphi''' \right) \, , \\
b_3 = & \, \tfrac{2}{9} \, r^2 \, \left( \varphi''_k + 2 r \varphi'''_k - \dot{\varphi}''_k \right) \, .
\end{split}
\ee

We now perform the Poisson resummation of the infinite sums. The functions $f_m$ entering the l.h.s.\ of the resummation
formula \eqref{prf} have the form $f_m = \theta\left(\tilde{\zeta}-\tfrac{r}{6} n^2 \right) \, (n^2)^m$ for $m=1,2,3$. Comparing this
structure with the general formula \eqref{prf}, it turns out to be convenient to set the free parameters to $x=0, a=1$. Defining $b \equiv \sqrt{6 \tilde{\zeta}/r}$
the Fourier transformed functions on the r.h.s.\ are of the general form
\be\label{intrep}
\tilde{f}_m(k) = b^{2m+1} \int_{-\infty}^\infty \diff x \, (x^2)^m \, \theta(1-x^2) \, e^{-ibkx} \, . 
\ee
The integrals are readily evaluated by noticing that
\be
\tilde{f}_0(k) = 2 k^{-1} \, \sin(b k) \, , 
\ee
which then serves as a generating function for the other values of $m$ by applying the recursion relation $\tilde{f}_m(k) = \left(- \p_k^2 \right)^m \, \tilde{f}_0(k)$. Upon
Poisson resummation, the flow equation \eqref{pdglr1} then reads
\be\label{pdglr2}
\begin{split}
\dot{\varphi}_k & +  3\varphi_k - 2r\varphi'_k =   \tfrac{r^{3/2}}{24 \sqrt{6} \pi^2} \, \sum_{n = - \infty}^\infty 
 \frac{b_1 \, \tilde{f}_1(2 \pi n) + b_2 \, \tilde{f}_2(2 \pi n) + b_3 \,\tilde{f}_3(2 \pi n)}{3\varphi_k+4(1-r-\ebb)\varphi'_k + 4\left(2-r-2\ebb\right)^2\varphi''_k} \,  \, . 
\end{split}
\ee
At this stage, it is illustrative to split the sum into its $n=0$-part and remainder. Realizing that $n$ only appears in even powers, the sum can be restricted to positive $n$. 
For $n = 0$ \eqref{intrep} directly leads to
\be
\tilde{f}_m(0) = \frac{2 }{2m+1} \, b^{2m+1} \, . 
\ee
The remaining sums $s_m \equiv \sum_{n=1}^\infty \tilde{f}_m(2\pi n)$ can be expressed in terms of polylogarithmic functions with argument $q^2 = e^{2 \pi i b}$
\be
\begin{split}
s_1 = & \tfrac{b^2}{2\pi} \left( i \ln(1-q^2) + \tfrac{1}{b\pi} {\rm Li}_2 + \tfrac{i}{2b^2\pi^2} {\rm Li}_3 + c.c. \right)  \, , \\
s_2 = & \tfrac{b^4}{2\pi} \left( i \ln(1-q^2) + \tfrac{2}{b\pi} {\rm Li}_2 + \tfrac{3i}{b^2\pi^2} {\rm Li}_3 - \tfrac{3}{b^3\pi^3} {\rm Li}_4 - \tfrac{3i}{2b^4\pi^4} {\rm Li}_5  + c.c. \right)  \, , \\
s_3 = & \tfrac{b^6}{2\pi} \left( i \ln(1-q^2) + \tfrac{3}{b\pi} {\rm Li}_2 + \tfrac{15i}{2b^2\pi^2} {\rm Li}_3 - \tfrac{15}{b^3\pi^3} {\rm Li}_4 - \tfrac{45i}{2b^4\pi^4} {\rm Li}_5 
+\tfrac{45}{2b^5\pi^5} {\rm Li}_6 +\tfrac{45i}{4b^6\pi^6} {\rm Li}_7 + c.c. \right)  \, ,
\end{split}
\ee
Here $c.c.$ denotes the complex conjugate so that the $s_m$ are real. In terms of this notation we arrive at the final form of the resummed flow equation
\be\label{pdglr}
\begin{split}
\dot{\varphi}_k  +  3\varphi_k - 2r\varphi'_k = & \, \frac{\tilde{\zeta}^{3/2}}{2 \pi^2} \, \frac{b_1  + b_2 \, b^2 + b_3 \, b^4}{3\varphi_k+4(1-r-\ebb)\varphi'_k + 4\left(2-r-2\ebb\right)^2\varphi''_k} \\
 & \, +  \frac{r^{3/2}}{12 \sqrt{6} \pi^2} \,  
 \frac{b_1 \, s_1 + b_2 \,s_2 + b_3 \, s_3}{3\varphi_k+4(1-r-\ebb)\varphi'_k + 4\left(2-r-2\ebb\right)^2\varphi''_k} \,  \, . 
\end{split}
\ee

At this stage some remarks on the structure of \eqref{pdglr} are in order. The first line of the r.h.s.\ actually encodes
the contributions of the \emph{local heat-kernel} and has an analytic expansion at $r = 0$. For $\ebb = 0$
this result agrees with the flow equation constructed in \cite{Demmel:2012ub}. This ``local part'' is dressed up by the non-local contributions
originating from the compact topology of the background three-sphere collected in the second line. Here $r$ appears in a non-analytic way and
$q^2$ has an essential singularity at $r=0$. One can also check explicitly that the resummed flow equation reproduces
the step-function behavior found in \eqref{pdglS3} through the branch cuts of the (poly-)logarithmic functions. While
expressing the sums in terms of polylogarithms may be beneficial for a deeper understanding of structural aspects, 
the numerical analysis of the infinite sums is more tractable, so that we stick to the flow equation \eqref{pdglS3} given
in the main part of the paper.

\section{Non-smooth functions and the heat kernel}
\label{App.expand}
In this appendix we want to clarify some possible caveat that appears when a functional trace such as
\begin{eqnarray}\label{simpleWtrace}
 {\rm Tr}\, W\!\left(\Box\right)
\end{eqnarray}
involves a non-smooth function $W\left(z\right)$.
For the sake of the argument in this appendix we will restrict our attention to the simple case in which
\begin{eqnarray} \label{simpleW}
 W\!\left(z\right) &\equiv& \theta \! \left(k^2-z\right)\,,
\end{eqnarray}
that captures the stepwise discontinuity of the function (\ref{eq:ourway}) appearing in the paper.
We simplify the computation by requiring $\Box\equiv\Delta$
to be the three-sphere Laplacian \cite{Rubin:1984tc}
and concentrate only on the local contributions to the heat kernel,
so that
\begin{eqnarray} \label{localheatkernel}
 K\left(s;x,x\right) &=& \left(4\pi s\right)^{-3/2}{\rm e}^{\frac{1}{6}Rs}\,.
\end{eqnarray}
The discussions of this section will extend trivially to the general case where winding modes are taken into account for the return probability of the heat kernel \cite{Camporesi:1990wm,Camporesi:1994ga}.

In the paper we extensively used the natural heat kernel definition
\begin{eqnarray} \label{simpleWtrace1}
 {\rm Tr}\, W\left(\Box\right)
 &\equiv&
 \int \!{\rm d}^3x \sqrt{g} \int \!{\rm d}s \, \widetilde{W}\!\left(s\right) K\left(s;x,x\right)\,
\end{eqnarray}
of the trace of a function $W\!\left(x\right)$ of $\Box$,
where we introduced $\widetilde{W}\!\left(s\right)$ as the inverse Laplace transform of $W\!\left(x\right)$.
When $W\!\left(x\right)$ does not admit an inverse Laplace transform,
the procedure has to be understood as over the limit of a sequence of smooth functions
that tends to $W\!\left(x\right)$ uniformly. Such a sequence exists for the particular case (\ref{simpleW}),
as well as for all the functions used through the paper.
We are thus free to formally manipulate (\ref{simpleWtrace1}) using the properties of the inverse Laplace transform, as well as the explicit form of the heat kernel (\ref{localheatkernel}).
The result is obtained by integrating over the argument $s$
\begin{eqnarray}\label{simpleWtrace2}
 {\rm Tr}\, W\!\left(\Box\right)
 &=&
 \frac{1}{\left(4\pi\right)^{3/2}} \int {\rm d}^3x \sqrt{g}\, Q_{3/2}\left(W\!\left(z-R/6\right)\right)\,,
\end{eqnarray}
where we used the definition of Mellin transform as introduced in the article (\ref{eq:mellin}). Evaluating it explicitly by means of (\ref{simpleW}) we obtain
\begin{eqnarray}\label{simpleWtrace3}
 {\rm Tr}\, W\!\left(\Box\right)
 &=& \frac{\left(6 k^2+R\right)^{3/2}}{18 \sqrt{6} \pi ^2}\, \theta\!\left(6 k^2+R\right)
\,.
\end{eqnarray}
The presence of the theta-function $\theta\!\left(6 k^2+R\right)$ is obviously a direct result of the discontinuity of $W\!\left(z\right)$ and cures the result from the presence of the branch-cut of the prefactor
$\left(6 k^2+R\right)^{3/2}$.

Let us now compute the trace (\ref{simpleWtrace}) by means of two different local expansions in $R$, which we will later resum. In the first case we expand the local heat kernel (\ref{localheatkernel})
\begin{eqnarray}
 K\left(s;x,x\right) &=&
 \left(4\pi s\right)^{-3/2}\sum_{n\ge 0}\frac{1}{n!}\left(\frac{s R}{6}\right)^n\,,
\end{eqnarray}
and use the expansion in (\ref{simpleWtrace1}) to obtain
\begin{eqnarray} \label{simpleWtrace4}
 {\rm Tr}\, W\!\left(\Box\right)
 &=&
 \frac{1}{\left(4\pi\right)^{3/2}}\int {\rm d}^3x \sqrt{g}
 \sum_{n\ge 0}\frac{1}{n!}\left(\frac{R}{6}\right)^n Q_{3/2-n}(W\!\left(z\right))
 \,.
\end{eqnarray}
Alternatively, it is possible to expand (\ref{simpleWtrace2}) directly inside the functional argument and use the linearity of the Mellin transforms in their argument to obtain
\begin{eqnarray} \label{simpleWtrace5}
 {\rm Tr}\, W\!\left(\Box\right)
 &=&
 \frac{1}{\left(4\pi\right)^{3/2}}\int {\rm d}^3x \sqrt{g}
 \sum_{n\ge 0}\frac{1}{n!}\left(-\frac{R}{6}\right)^n Q_{3/2}(W^{(n)}\!\left(z\right))
 \,,
\end{eqnarray}
where we denote by $W^{(n)}\!\left(z\right)$ the $n$-th derivative of $W\!\left(z\right)$.
Using the explicit form (\ref{simpleW}), it is not hard to show that, for any $k>0$
\begin{eqnarray}
 Q_{3/2}(W^{(n)}\!\left(z\right)) &=& \left(-1\right)^n Q_{3/2-n}(W\!\left(z\right))\,,
\end{eqnarray}
that implies the equivalence of (\ref{simpleWtrace4}) and (\ref{simpleWtrace5}) at almost all scales $k$,
though differences may arise when taking the $k\to 0$ limit corresponding to the infrared.

We elaborate further resumming the two series (\ref{simpleWtrace4}) and (\ref{simpleWtrace5}) in the case $k>0$.
The most convenient way to proceed is to compute
\begin{eqnarray}
 Q_{3/2-n}(W\!\left(z\right)) &=& \frac{k^{3-2 n}}{2 \pi ^{3/2} (3-2 n) \Gamma \left(\frac{3}{2}-n\right)}\,,
\end{eqnarray}
which can be obtained only by continuing the result in $n$ from the range $n<3/2$.
The series (\ref{simpleWtrace4}) and (\ref{simpleWtrace5}) can now be resummed to
\begin{eqnarray} \label{simpleWtrace6}
 {\rm Tr}\, W\!\left(\Box\right)
 =
 \frac{1}{\left(4\pi\right)^{3/2}}\int {\rm d}^3x \sqrt{g}
 \sum_{n\ge 0}
 \frac{4 k^{3-2 n}R^n}{(3-2 n) 6^n \Gamma \left(\frac{3}{2}-n\right)n!}
 =
 \frac{\left(6 k^2+R\right)^{3/2}}{18 \sqrt{6} \pi ^2}
 \,.
\end{eqnarray}
The resummation (\ref{simpleWtrace6}) displays the branch cut of the prefactor of (\ref{simpleWtrace3}),
but this time it is not cured by the presence of a step function.
It is easy to understand why by directly expanding (\ref{simpleWtrace3}) around $R=0$. The step function will contribute to the $n$-th order of the expansion with derivatives of the delta-function 
$\delta^{(m)}\left(k^2\right)$ for $m<n$, which are zero for any $k>0$. We therefore conclude that (\ref{simpleWtrace3}) and (\ref{simpleWtrace6}) admit the same local expansion in $R$.

As a remark, if the same procedure is carried out in an even dimensional sphere, the result (\ref{simpleWtrace6}) is not expected to display any branch-cut due to the fact that the corresponding prefactor would have an integer power \cite{Benedetti:2013nya}.
Apparently then the even dimensional case can be continued easily below any desired value.
However, it should be clear from the investigation above that any analytic continuation of the resummed local expansion in any dimensionality has to be taken with care since it would otherwise fail to correctly take care of branch cuts such as those appearing in (\ref{simpleWtrace3}).
The investigations of our paper need a computation of (\ref{simpleWtrace}) that is valid for any value of $R$ and $k$, and the only viable option in this direction is the method (\ref{simpleWtrace2}) that lead to (\ref{simpleWtrace3}) and that we applied through the paper.

\end{appendix}
%

\end{document}